\documentclass[12pt]{article}
\usepackage[a4paper, total={7in, 10in}]{geometry}
\usepackage[parfill]{parskip}
\usepackage{physics, tensor, float, subcaption}
\usepackage{graphicx}
\graphicspath{ {Plots/} }
\usepackage{jhep-mod}
\usepackage{bm}
\usepackage{soul}
\usepackage{amssymb,amsmath,amsthm}
\usepackage{mathrsfs}
\usepackage[utf8]{inputenc}
\usepackage{enumerate}
\usepackage{bigints}
\usepackage{xcolor}
\usepackage{appendix}
\usepackage{graphicx}
\usepackage{float}
\usepackage{tikz}
\usepackage{setspace}
\usepackage{cancel}
\usepackage{doi}
\definecolor{purple}{rgb}{1,0,1}
\newcommand{\blue}[1]{{\color{blue} #1}}

\definecolor{lime}{HTML}{A6CE39}
\newcommand{\orcidicon}{%
	\begin{tikzpicture}
	\draw[lime, fill=lime] (0,0) 
		circle [radius=0.16] 
		node[white] {{\fontfamily{qag}\selectfont \tiny ID}};
	\draw[white, fill=white] (-0.0625,0.095) 
		circle [radius=0.007];
	\end{tikzpicture}
	\hspace{-5mm}
}
\newcommand\orcidJosh{{\href{https://orcid.org/0000-0003-1200-7261}{\orcidicon}}}
\newcommand\orcidThomas{{\href{https://orcid.org/0000-0002-0314-4136}{\orcidicon}}}
\newcommand\orcidAlex{{\href{https://orcid.org/0000-0002-1763-3563}{\orcidicon}}}
\newcommand\orcidMatt{{\href{https://orcid.org/0000-0003-1088-6485}{\orcidicon}}}
\begin{document}
%========================================================
%========================================================
%========================================================

\title{\vspace{-25pt}\huge{
Geodesics for the Painlev\'e--Gullstrand form of Lense--Thirring spacetime
}}

%========================================================
%========================================================
%========================================================
\author{
\Large
Joshua Baines\!\orcidJosh$^1$, Thomas Berry\!\orcidThomas$^2$, 
\\
Alex Simpson\!\orcidAlex$^1$,\! 
{\sf  and} Matt Visser\!\orcidMatt$^1$}
%========================================================
%========================================================
%========================================================
%========================================================
\affiliation{
$^1$ School of Mathematics and Statistics, Victoria University of Wellington, 
\\
\null\qquad PO Box 600, Wellington 6140, New Zealand.}
\affiliation{
$^2$ Robinson Institute,
Victoria University of Wellington, 
\\
\null\qquad PO Box 600, Wellington 6140, New Zealand.}
%========================================================
%========================================================
\emailAdd{joshua.baines@sms.vuw.ac.nz}
\emailAdd{thomas.berry@vuw.ac.nz}
\emailAdd{alex.simpson@sms.vuw.ac.nz}
\emailAdd{matt.visser@sms.vuw.ac.nz}
%========================================================
%========================================================

\abstract{
\vspace{1em}

Recently, the current authors have formulated and extensively explored a rather novel Painlev\'{e}--\-Gullstrand variant of the slow-rotation Lense--Thirring spacetime, a variant which has particularly elegant features --- including unit lapse, intrinsically flat spatial 3-slices, and a separable Klein--Gordon equation (wave operator). This spacetime also possesses 
a non-trivial  Killing tensor, implying separability of the Hamilton--Jacobi equation, the existence of a Carter constant,  and complete \emph{formal} integrability of the the geodesic equations. Herein we investigate the geodesics in some detail, in the general situation demonstrating the occurrence of ``ultra-elliptic'' integrals. Only in certain special cases can the complete geodesic integrability be explicitly cast in terms of elementary functions. The model is potentially of astrophysical interest both in the asymptotic large-distance limit and as an example of a ``black hole mimic'', a controlled deformation of the Kerr spacetime that can be contrasted with ongoing astronomical observations.

\bigskip
\noindent
{\sc Date:} Friday 10 December  2021; \LaTeX-ed \today


\bigskip
\noindent{\sc PhySH:} 
Gravitation
}

%========================================================
\maketitle
%========================================================
\def\tr{{\mathrm{tr}}}
\def\diag{{\mathrm{diag}}}
\def\cof{{\mathrm{cof}}}
\def\pdet{{\mathrm{pdet}}}
\def\d{{\mathrm{d}}}
\def\K{{\mathcal{K}}}
\def\O{{\mathcal{O}}}
\parindent0pt
\parskip7pt
\newcommand{\C}{\mathcal{C}}

\clearpage
%=====================================================
\section{Introduction}
\label{S:intro}
%=====================================================

Recently the current authors have introduced and extensively explored a specific new variant of the slow-rotation Lense--Thirring spacetime~\cite{4-of-us,PGLT2}, described by the explicit line element
\begin{equation}
\label{E:pglt}
\d s^2 = - \d t^2 +\left\{\d r+\sqrt{\frac{2m}{r}} \; \d t\right\}^2
+ r^2 \left\{\d\theta^2+\sin^2\theta\; \left(\d\phi - {2J\over r^3} \d t\right) ^2\right\} \ .
\end{equation}
For this variant of the Lense--Thirring spacetime the metric possesses both unit lapse~\cite{Unit-lapse}, and also exhibits a flat spatial $3$-metric~\cite{4-of-us,PGLT2}. That is, the spacetime metric is presented in so-called Painlev\'{e}--Gullstrand form~\cite{painleve1,painleve2,gullstrand,poisson}, with a relatively simple globally defined tetrad~\cite{4-of-us,PGLT2}. 
These purely mathematical observations make this spacetime of particular theoretical interest~\cite{Rajan:2016, Skakala:2010}.

We emphasize that there is no Birkhoff-like theorem for axi\-symmetric spacetimes in $(3+1)$ dimensions~\cite{Birkhoff, Jebsen, Deser, Ravndal, Martinez:2004, Skakala}. The Kerr solution need not, (and typically will not), perfectly model rotating horizonless astrophysical sources such as stars and planets, due to the nontrivial mass multipole moments that such objects typically possess. Instead, the Kerr solution will only model the gravitational field in the asymptotic large-distance regime, a region where the Lense--Thirring spacetime serves as a perfectly valid approximation to Kerr~\cite{Lense-Thirring, 
Lense-Thirring-translation,
Pfister, Adler-Bazin-Schiffer, MTW, Wald, weinberg, Hobson, D'Inverno, Hartle, Carroll, kerr-intro, kerr-book, Kerr, Kerr-Texas, kerr-newman, race, kerr-book-2, river, Doran, Gordon-form}. Given that this variant of the Lense--Thirring metric is a valid approximation for the gravitational fields of rotating stars and planets in the same regime that the Kerr solution is appropriate, there is a compelling physics argument to use the Painlev\'{e}--Gullstrand form of Lense--Thirring to model various astrophysically interesting cases~\cite{Carballo-Rubio:2018,Visser:2009a,Visser:2009b, Vincent:2012, Vincent:2020, Bambi:2012, Glampedakis:2018}. 

\enlargethispage{40pt}
From a purely theoretical perspective, 
the Lense--Thirring metric is algebraically \emph{much} simpler than the Kerr metric, making most calculations  significantly easier to conduct, and the Lense--Thirring metric can be recast into Painlev\'{e}--Gullstrand form, while the Kerr metric cannot~\cite{Kerr-Darboux,Valiente-Kroon:2004a,Valiente-Kroon:2004b,Jaramillo:2007}. This spacetime exhibits a separable Klein--Gordon equation (wave operator)~\cite{PGLT2} and also possesses a non-trivial Killing tensor, thereby implying separability and complete (formal) integrability of the Hamilton--Jacobi equations for geodesic motion~\cite{PGLT2}. 
Below, we shall discuss two particularly interesting classes of geodesics; the generic case involving ultra-elliptic integrals, and the case of vanishing Carter constant where the analysis can be completely performed in terms of elementary functions. This should be compared to what can and cannot be done for the usual Kerr spacetime~\cite{Hioki:2009, Wilkins:1972, Page:2006, Pretorius:2007, Fujita:2009, Hackmann:2010, Sereno:2006, Sereno:2007, Gralla:2019-a, Gralla:2019-b, Warburton:2013, deFelice:1968, Rana:2019, Gariel:2007, Gariel:2016, Paganini:2016, Lammerzahl:2015, Boccaletti:2005}.

Observationally, apart from its interest in the large-distance asymptotic regime, this Lense--Thirring variant may also be viewed as a ``black hole mimic'' that can be contrasted with ongoing astronomical observations of various black hole candidates~\cite{Carballo-Rubio:2018, LISA, Echos, bb-Kerr, bb-KN, eye-of-storm}.

\clearpage
We note that a competing slow-rotation model has recently been discussed in reference~\cite{Finn}. The trade-off made therein was to improve the integrability properties (the ``hidden symmetries") at the cost of sacrificing the global Painlev\'e--Gullstrand form of the metric. 

%==============================
\section{Killing Tensor and Carter constant}
\label{S:Killing}
%==============================
\enlargethispage{20pt}
Based on the algorithm presented in two recent papers by Papadopoulos and Kokkotas~\cite{Papadopoulos:2018, Papadopoulos:2020}, which are in turn based on considerably older results by Benenti and Francaviglia~\cite{Benenti:1979},  we found the non-trivial Killing tensor~\cite{PGLT2}:
\begin{equation}
K_{ab} \; \d x^a \; \d x^b = 
r^4\left\{\d\theta^2 + \sin^2\theta \left(\d\phi - {2J\over r^3} \d t\right)^2 \right\}.
\end{equation}
Explicitly, we write the metric as~\cite{4-of-us,PGLT2}:
\begin{equation}
g_{ab} = \left[ \begin{array}{c|ccc}
-1+{2m\over r} + {4 J^2\sin^2\theta\over r^4} &  \sqrt{2m\over r} & 0 & -{2J\sin^2\theta\over r}\\
\hline
\sqrt{2m\over r} & 1 & 0 & 0\\
0& 0 & r^2 & 0\\
-{2J\sin^2\theta\over r} & 0 & 0 & r^2\sin^2\theta\\
\end{array}
\right]_{ab}.
\end{equation}
Then $\det(g_{ab}) = - r^4 \sin^2\theta$, and for the inverse metric we have~\cite{4-of-us,PGLT2}:
\begin{equation}
\label{E:contra_metric}
g^{ab} = \left[ \begin{array}{c|ccc}
-1&  \sqrt{2m\over r} & 0 & -{2J\over r^3}\\
\hline
\sqrt{2m\over r} & 1-{2m\over r} & 0 & \sqrt{2m\over r} \;{2J\over r^3}\\
0& 0 &\; {1\over r^2}\; & 0\\
-{2J\over r^3} &  \sqrt{2m\over r}\; {2J\over r^3} & 0 & {1\over r^2\sin^2\theta} - {4J^2\over r^6}\\
\end{array}
\right]^{ab}.
\end{equation}
The (contravariant) non-trivial Killing tensor is~\cite{PGLT2}:
\begin{equation}
K^{ab}=
\begin{bmatrix}
0& 0 & 0 & 0\\
0 & 0 & 0 & 0\\
0 & 0 &\; 1 \;& 0\\
0 & 0 & 0 & {1\over\sin^2\theta}
\end{bmatrix}
^{ab} \ .
\end{equation}
The corresponding covariant form of the Killing tensor,  
$K_{ab}= g_{ac} \,K^{cd}\, g_{db}$, is then~\cite{PGLT2}:
\begin{equation}
\label{PGLT_Killing_Tensor}
K_{ab}=
\begin{bmatrix}
\frac{4J^2\sin^2\theta}{r^2} & 0 & 0 & -2Jr\sin^2\theta\\
0 & 0 & 0 & 0\\
0 & 0 & \; r^4\;  & 0\\
-2Jr\sin^2\theta & 0 & 0 & r^4\sin^2\theta
\end{bmatrix}
_{ab} \ .
\end{equation}
One can easily explicitly check ({\emph{e.g.}}, {\sf{Maple}}) that $\nabla_{(c}K_{ab)}= K_{(ab;c)} = 0$.

For any affine parameter $\lambda$, the (generalized) Carter constant is now~\cite{PGLT2}:
\begin{equation} 
\label{E:Carter0}
\mathcal{C}=K_{ab}\;\frac{\d x^a}{\d\lambda}\,\frac{\d x^b}{\d\lambda}=r^4\left[ \left(\frac{\d\theta}{\d\lambda}\right)^2 + \sin^2\theta \left(\frac{\d\phi}{\d\lambda}-\frac{2J}{r^3}\frac{\d t}{\d\lambda}\right)^2 \right] \ ,
\end{equation}
Without any loss of generality we may choose $\lambda$ be future-directed, (so $\d\lambda/\d t > 0$). Note that by construction, since it is a sum of squares,  $\C \geq 0$.
(For additional recent discussion on general Killing tensors see~\cite{Frolov:2017, Giorgi}.) 

%==============================
\section{Four conserved quantities}
\label{S:four}
%==============================

In addition to the Carter constant (\ref{E:Carter0}), in this spacetime geometry we have three other conserved quantities. Two of these come from the time-translation and axial Killing vectors,  $\xi^a=(1;0,0,0)^a$ and  $\psi^a=(0,0,0,1)^a$, respectively:
\begin{equation}\label{E}
E = -\xi_a\dfrac{\d x^a}{\d\lambda} = \left(1-\frac{2m}{r}-\frac{4J^2\sin^2\theta}{r^4} \right)\frac{\d t}{\d\lambda}-\sqrt{\frac{2m}{r}}\frac{\d r}{\d\lambda}+\frac{2J\sin^2\theta}{r}\frac{\d\phi}{\d\lambda} \ ;
\end{equation}
and
\begin{equation} \label{L}
L = \psi_a\frac{\d x^a}{\d\lambda} = r^2\sin^2\theta\frac{\d\phi}{\d\lambda} -\frac{2J\sin^2\theta}{r}\frac{\d t}{\d\lambda} \ .
\end{equation}
The final conserved quantity, $\epsilon$, 
with $\epsilon \in \{0,-1\}$ for null and timelike geodesics respectively, 
comes from the trivial Killing tensor (the metric $g_{ab}$):
\begin{equation}\label{Geo_eqn}
\begin{split}
\epsilon=g_{ab}\frac{\d x^a}{\d\lambda}\frac{\d x^b}{\d\lambda}= & -\left(\frac{\d t}{\d\lambda}\right)^2+\left(\frac{\d r}{\d\lambda}+\sqrt{\frac{2m}{r}}\frac{\d t}{\d\lambda}\right)^2\\
& + r^2\left[\left(\frac{\d\theta}{\d\lambda}\right)^2+\sin^2\theta\left(\frac{\d\phi}{\d\lambda}-\frac{2J}{r^3}\frac{\d t}{\d\lambda}\right)^2\right] \ .
\end{split}
\end{equation}

%\enlargethispage{30pt}
We can greatly simplify these four conserved quantities by rewriting them as:
\begin{equation} \label{L_2}
L=r^2\sin^2\theta \left(\frac{\d\phi}{\d\lambda}-\frac{2J}{r^3}\frac{\d t}{\d\lambda}\right) \ ;
\end{equation}
\begin{equation} \label{C_2}
\mathcal{C}=r^4\left(\frac{\d\theta}{\d\lambda}\right)^2 +{L^2\over \sin^2\theta} \ ;
\end{equation}
\begin{equation}\label{Geo_eqn_2}
\epsilon=  -\left(\frac{\d t}{\d\lambda}\right)^2+\left(\frac{\d r}{\d\lambda}+\sqrt{\frac{2m}{r}}\frac{\d t}{\d\lambda}\right)^2 +{\mathcal{C}\over r^2} \ ;
\end{equation}
\begin{equation}\label{E_2}
E=\left(1-\frac{2m}{r}\right)\frac{\d t}{\d\lambda} - \sqrt{\frac{2m}{r}}\frac{\d r}{\d\lambda} + \frac{2J}{r^3}L \ .
\end{equation}
Notice that by construction $\C \geq L^2\geq 0$, and that $(\d t/\d\lambda)^2+\epsilon \geq 0$.

If $\epsilon=0$ then, without  loss of generality, we can rescale the affine parameter $\lambda$ to set \emph{one} of the constants $\{\C,E,L\} \to 1$. It is perhaps most intuitive to set $E\to 1$.

In contrast if $\epsilon=-1$ then $\lambda = \tau$ is the proper time and there is no further freedom to rescale the affine parameter. $E$ then has real physical meaning and the qualitative behaviour is governed by the \emph{sign} of $E^2+\epsilon$. Concretely, at least in the case of Carter constant zero, one asks: 
\begin{itemize}
\item 
Is $E <1$? (Bound orbits.) 
\item
Is $E=1$? (Marginal orbits.) 
\item 
Or is $E>1$? (Unbound orbits).
\end{itemize}

\paragraph{Forbidden declination range:}
The form of the Carter constant, equation \eqref{C_2}, 
since it is a sum of squares, gives a range of forbidden declination angles for any given, non-zero values of $\mathcal{C}$ and $L$. 
We require that $\d\theta/\d\lambda$ be real, and from equation \eqref{C_2} this implies the following requirement:
\begin{equation}
\left(r^2\frac{\d\theta}{\d\lambda}\right)^2=\mathcal{C}-\frac{L^2}{\sin^2\theta}\geq 0
\quad\Longrightarrow\quad
\sin^2 \theta \geq {L^2\over \C} \ .
\end{equation}
Then provided $\C \geq L^2$, which is automatic in view of \eqref{C_2}, we can define a critical angle $\theta_* \in [0,\pi/2]$ by setting 
\begin{equation}
\theta_* = \sin^{-1}(|L|/\sqrt{\C}) \ .
\end{equation}
 Then the allowed range for $\theta$ is the equatorial band:
\begin{equation}
\label{E:theta_range}
\theta \in \Big[ \theta_*, \pi -\theta_*\Big] \ .
\end{equation}
\begin{itemize}
\item 
For $L^2=\C$ we have $\theta=\pi/2$; the motion is restricted to the equatorial plane.
\item
For $L=0$ with $\C>0$ the range of $\theta$ is \emph{a priori} unconstrained; $\theta\in[0,\pi]$.
\item
For $L=0$ with $\C=0$ the declination is fixed, $\theta(\lambda)=\theta_{0}$, and the motion is restricted to a constant declination conical surface.
\end{itemize}

\enlargethispage{40pt}
%==============================
\section{General geodesics}
\label{S:generalgeo}
%==============================

\noindent 
Using equations \eqref{L_2}, \eqref{C_2}, \eqref{Geo_eqn_2} \& \eqref{E_2} we can explicitly and analytically solve for the four unknown functions $\d t/\d\lambda$, $\d r/\d\lambda$, $\d\theta/\d\lambda$ and $\d\phi/\d\lambda$ as explicit functions of $r$ and $\theta$, parameterized by the four conserved quantities $\C$, $E$, $L$, and $\epsilon$, as well as the quantities $m$ and $J$ characterizing mass and angular momentum of the central object. 

In solving for these unknown functions, one encounters three independent sign choices. (Note $\d t/\d\lambda$ is always taken to be positive.) 
Consequently, it is useful to define the following quantities (the subsequent physical interpretations are chosen from context for each equation):
\begin{eqnarray}
    S_r  &=& \left\{
    \begin{array}{rl}
    +1 & \qquad\mbox{outgoing geodesic} \\
     -1 & \qquad\mbox{ingoing geodesic}
    \end{array}\right. \ ; \label{Sphi}\\
    && \nonumber \\
    S_{\theta} &=& \left\{
    \begin{array}{rl}
    +1 & \qquad\mbox{increasing declination geodesic} \\
     -1 & \qquad\mbox{decreasing declination geodesic}
    \end{array}\right. \ ; \\
    && \nonumber \\
    S_{\phi} &=& \left\{
    \begin{array}{rl}
    +1 & \qquad\mbox{prograde geodesic} \\
     -1 & \qquad\mbox{retrograde geodesic}
    \end{array}\right. \ .
\end{eqnarray}

%\clearpage
%==============================
\subsection{Trajectories}
\label{SS:trajectories}
%==============================

For the geodesic trajectories we find the four equations:
\begin{equation}
\frac{\d r}{\d\lambda}= S_r \sqrt{X(r)} \ ;
\label{E:r}
\end{equation}
\begin{equation}
\frac{\d t}{\d\lambda}= \frac{E-2JL/r^3+ S_r \sqrt{(2m/r)X(r)}}{(1-2m/r)} \ ;
\label{E:t}
\end{equation}
\begin{equation}
\frac{\d\theta}{\d\lambda}= S_{\theta}\frac{\sqrt{\mathcal{C}-L^2/\sin^2\theta}}{r^2} \ ;
\label{E:theta}
\end{equation}
%and
\begin{equation}
\frac{\d\phi}{\d\lambda}= \frac{L}{r^2\sin^2\theta} + 2J\; \frac{E-2JL/r^3 + S_{\phi}\sqrt{(2m/r)X(r)}}{r^3(1-2m/r)} \ . \label{E:phi}
\end{equation}
Here $X(r)$ is the sextic Laurent polynomial
\begin{equation} \label{X1}
X(r) =\left(E - \frac{2JL}{r^3}\right)^2-\left(1-\frac{2m}{r}\right)\left(-\epsilon + \frac{\mathcal{C}}{r^2}\right) \ .
\end{equation}
We note
\begin{equation}
\lim_{r\to\infty} X(r) = E^2+\epsilon \ .
\end{equation}
In terms of the roots of this polynomial in the generic case we can write
\begin{equation}
X(r) = {E^2+\epsilon\over r^6} \;\prod_{i=1}^6 (r-r_i) \ .
\end{equation}\enlargethispage{20pt}
In the special case $E^2+\epsilon=0$, corresponding to a marginally bound timelike geodesic, the sextic degenerates to a quintic
\begin{equation}
X(r) = {2m\over r^5} \; \prod_{i=1}^5 (r-r_i) \ .
\end{equation}

Qualitatively the radial motion can be bounded (if one is trapped between two real roots), or diverge to spatial infinity (if one is trapped above the outermost real root), or be a plunge to $r=0$ (if one is trapped below the innermost positive real root). In the immediate vicinity of any real root $r_i$ the behaviour will depend on the multiplicity $m_i$ of that root. Approximately one has
\begin{equation}
\frac{\d r}{\d\lambda}\approx\pm 2 K_i \sqrt{|r-r_i|^{m_i}} \ .
\end{equation}
If the root is multiplicity $1$, (the generic situation) one has a ``bounce'' at the turning point (perinegricon or aponegricon --- periapsis or apoapsis) at some finite value of the affine parameter:
\begin{equation}
|r-r_i| \approx   K_i ^2(\lambda-\lambda_0)^2 \ .
\end{equation}
If the root is multiplicity $2$, (a somewhat rarer situation), one has an exponential approach to the root:
\begin{equation}
|r-r_i| \approx  C_i \exp(\pm 2 K_i \; \lambda).
\end{equation}\enlargethispage{20pt}
If the root is multiplicity $3$ or higher, (an unusual situation; for instance take $\epsilon\to0$ and $\C\to 0$),  one has a very slow polynomial approach to the root as $|\lambda| \to \infty$:
\begin{equation}
|r-r_i| \approx  \Big| K_i (m_{i}-2) (\lambda-\lambda_0) \Big|^{2\over 2-m_i} \ .
\end{equation}

%==============================
\subsection{Integrating the affine parameter}
\label{SS:affine}
%==============================

From equation \eqref{E:r}, we find
\begin{equation} \label{lambda_r}
\lambda(r) = \lambda_0 + S_r  \int_{r_0}^r {\d\bar r\over\sqrt{X(\bar r)}} 
= \lambda_0 + \int_{r_0}^r {|\d\bar r|\over\sqrt{X(\bar r)}} \ .
\end{equation}
This can also be re-expressed as follows:
\begin{itemize}
    \item Outgoing geodesic $\Longrightarrow S_r=+1 \ \cap \ r_0<r$, hence $\lambda(r) = \lambda_0 + \int_{r_0}^{r}{\d\bar r\over\sqrt{X(\bar r)}}$;
    \item Ingoing geodesic $\Longrightarrow S_r=-1 \ \cap \ r_0>r$, hence $\lambda(r) = \lambda_0 + \int_{r}^{r_0}{\d\bar r\over\sqrt{X(\bar r)}}$.
\end{itemize}
Integrals of this type are known as ultra-elliptic integrals~\cite{hyper,hyper2}, and date back (at least) to work by Weierstrass and Kovalevskaya in the second half of the 19$^{th}$ century.

%\clearpage
If $X(r)$ were to be cubic or quartic, this would be an ordinary elliptic integral~\cite{elliptic}. If $X(r)$ is quintic or sextic, as above, this is an ultra-elliptic integral. More generally, for polynomials of arbitrary order, these would be called hyper-elliptic integrals. Even more generally, these integrals are a sub-class of the so-called Abelian integrals~\cite{abelian}.
Generically, equation (\ref{lambda_r}) cannot be explicitly integrated in closed form using only elementary functions, hence we cannot analytically invert this relation to find $r(\lambda)$. 
However there is no obstacle in principle to numerical integration to explicitly find the affine parameter $\lambda(r)$. (Even for the exact Kerr solution one rapidly finds that use of some level of numerical integration is almost unavoidable~\cite{Yang:2013-a, Yang:2013-b, Chan:2017}.)

Note that if one is trapped (above or below) any real root of the polynomial $X(r)$ of multiplicity $1$, 
then every time one ``bounces'' off the root the quantity $S_r$ will flip sign, and $\lambda(r)$ will be double-valued though the inverse function $r(\lambda)$ will always be single-valued. This is as it should be to guarantee that the affine parameter $\lambda$ is always continuously increasing with time.

%\clearpage
\enlargethispage{30pt}
Note that if one is trapped between two real roots of the polynomial $X(r)$, say $r_{min}$ and $r_{max}$, both of multiplicity $1$, then each bounce from  $r_{min}$ to $r_{max}$, or from  $r_{max}$ to $r_{min}$, will advance the affine parameter by some finite amount (the appropriate ``period'' of the ultra-elliptic integral, now typically called a complete ultra-elliptic integral):
\begin{equation}
\Delta\lambda = \int_{r_{min}}^{r_{max}} {\d\bar r\over\sqrt{X(\bar r)}} \ .
\end{equation}
Then $\lambda(r)$ will be multi-valued though the inverse function $r(\lambda)$ will always be single-valued.  Furthermore, $r(\lambda)$, while analytically intractable,  will at least be known to be periodic, with known periodicity $2\,\Delta\lambda$. If instead one is approaching a real root of multiplicity 2 or higher, then $\lambda(r) $ will diverge --- one will not actually reach the root for any finite amount of affine parameter lapse --- that is $r(\lambda)$ will asymptote to that higher multiplicity root.

%==============================
\subsection{Integrating the epoch}
\label{SS:epoch}
%==============================

Using equations \eqref{E:r} \& \eqref{E:t}, we find
 \begin{equation}
     \frac{\d t}{\d r} = \frac{\sqrt{2mr}}{r-2m} + S_r \frac{E-2JL/r^3}{(1-2m/r)\sqrt{X(r)}} \ ,
 \end{equation}
so
\begin{equation}
t(r) = t_0 + \int_{r_0}^r \left(\frac{\sqrt{2m\bar{r}}}{\bar{r}-2m} + S_r\frac{E-2JL/\bar{r}^3}{(1-2m/\bar{r})\sqrt{X(\bar{r})}}\right) \d\bar r \ ,
\end{equation}
where now $S_r=\mbox{sign}(r-r_0)$.

It is straightforward to integrate the first term in the integrand, yielding
\begin{eqnarray}\label{E:t(r)}
    t(r) &=& t_0 + 2\sqrt{2m}(\sqrt{r}-\sqrt{r_0})+ 2m\ln\left[\frac{(\sqrt{r}-\sqrt{2m})(\sqrt{r_0}+\sqrt{2m})}{(\sqrt{r_0}-\sqrt{2m})(\sqrt{r}+\sqrt{2m})}\right] \nonumber \\
    && \nonumber \\
    && \qquad \qquad \qquad \qquad \qquad \qquad \quad + S_r \int_{r_0}^{r}\frac{E-2JL/\bar{r}^3}{(1-2m/\bar{r})\sqrt{X(\bar{r})}}\;\d\bar{r} \ . \label{tanh1}
\end{eqnarray}
As before, the remaining integral is an ultra-elliptic integral~\cite{hyper,hyper2}, now a \emph{different} ultra-elliptic integral, and this equation cannot be explicitly integrated in closed form.

%\enlargethispage{50pt}
Note that if one is trapped between two real roots of the polynomial $X(r)$, say $r_1$ and $r_2$, both of multiplicity $1$, and both outside the horizon at $r=2m$, then each bounce from  $r_1$ to $r_2$ will advance the Killing time by some finite amount
\begin{equation}
T(r_1,r_2) = \int_{r_{1}}^{r_{2}} \left(\frac{\sqrt{2m\bar{r}}}{\bar{r}-2m} + S_r\frac{E-2JL/\bar{r}^3}{(1-2m/\bar{r})\sqrt{X(\bar{r})}}\right) \d\bar r \ ,
\end{equation}
where now $S_r = \mathrm{sign}{(r_2-r_1)}$.

Specifically, if $r_{max} = \max\{r_1,r_2\}$ and $r_{min} = \min\{r_1,r_2\}$, then on the upswing from $r_{min}$ to $r_{max}$ one has
\begin{equation}
T_{up} = \int_{r_{min}}^{r_{max}} \left(\frac{\sqrt{2m\bar{r}}}{\bar{r}-2m} +\frac{E-2JL/\bar{r}^3}{(1-2m/\bar{r})\sqrt{X(\bar{r})}}\right) \d\bar r \ .
\end{equation}
In contrast on the downswing from $r_{max}$ to $r_{min}$ one has
\begin{equation}
T_{down} = \int_{r_{max}}^{r_{min}} \left(\frac{\sqrt{2m\bar{r}}}{\bar{r}-2m} -\frac{E-2JL/\bar{r}^3}{(1-2m/\bar{r})\sqrt{X(\bar{r})}}\right) \d\bar r \ .
\end{equation}
That is
\begin{equation}
T_{down} = \int_{r_{min}}^{r_{max}} \left(-\frac{\sqrt{2m\bar{r}}}{\bar{r}-2m} +\frac{E-2JL/\bar{r}^3}{(1-2m/\bar{r})\sqrt{X(\bar{r})}}\right) \d\bar r \ .
\end{equation}
The total period is therefore
\begin{equation}
T = T_{up}+T_{down} 
= 2\int_{r_{min}}^{r_{max}} \frac{E-2JL/\bar{r}^3}{(1-2m/\bar{r})\sqrt{X(\bar{r})}} \; \d\bar r \ .
\end{equation}
This is again a complete ultra-elliptic integral, now a \emph{different} complete ultra-elliptic integral.

In this situation $t(r)$ will be multi-valued while the inverse function $r(t)$ will be single valued.  While analytically intractable, $r(t)$ will at least be known to be periodic, with known periodicity $2\,T$. If instead one is approaching a real root of multiplicity 2 or higher, then $t(r) $ will diverge --- one will not actually reach the root for any finite amount of Killing time.

%================================
\subsection{Integrating the declination}
\label{SS:declination}
%================================

As for our equation involving the declination angle $\theta$, first recall the definition of the critical angle $\theta_*$ as $\theta_*=\sin^{-1}\left(\vert L\vert/\sqrt{\mathcal{C}}\right)$. Then from equation \eqref{E:theta}, we find
\begin{eqnarray}
\frac{\d\cos\theta}{\d\lambda} 
&=& -S_{\theta} {\sqrt{\mathcal{C}\sin^2 \theta-L^2}\over r^2} 
\nonumber\\
&=& -S_{\theta}\frac{\sqrt{\mathcal{C}}}{r^2}\sqrt{\sin^2\theta-\sin^2\theta_*}\;
\nonumber\\
&=& -S_{\theta} \frac{\sqrt{\mathcal{C}}}{r^2}\sqrt{\cos^2\theta_*-\cos^2\theta} \ ,
\end{eqnarray}
implying
\begin{equation}
{\d\cos\theta \over  \sqrt{\cos^2\theta_*-\cos^2\theta}}  = -S_{\theta}  \frac{\sqrt{\mathcal{C}}}{r^2}\,\d\lambda = \left(-S_{\theta}\frac{\sqrt{\mathcal{C}}}{r^2}\right)\left(S_r \frac{\d r}{\sqrt{X(r)}}\right) \ .
\end{equation}
From this we see
\begin{equation}
{\d\cos^{-1}\left(\frac{\cos\theta}{\cos\theta_*}\right)} = S_{\theta}S_r \frac{\sqrt{\C}}{r^2}\frac{\d r}{\sqrt{X(r)}} \ ,
\end{equation}
that is
\begin{equation}
\label{E:declination}
 {\cos^{-1}\left(\frac{\cos\theta}{\cos\theta_*} \right)} =  {\cos^{-1}\left(\frac{\cos\theta_0}{\cos\theta_*} \right)} + S_{\theta}S_r \sqrt{\mathcal{C}} \int_{r_0}^r\frac{\d\bar{r}}{\bar{r}^2\sqrt{X(\bar{r})}} \ .
\end{equation}
Without loss of generality we may allow the geodesic to reach the critical angle $\theta_*$ at some radius $r_*$, and then use that as our new initial data. 

This effectively sets $\theta_0=\theta_*$, and gives us the following simplified result:
\begin{equation}
\label{E:declination2}
    \cos^{-1}\left(\frac{\cos\theta}{\cos\theta_*}\right) = 
    S_{\theta}S_r \sqrt{\mathcal{C}}\int_{r_*}^{r}\frac{\d\bar{r}}{\bar{r}^2\sqrt{X(\bar{r})}} \ .
\end{equation}
Thence
\begin{eqnarray}
\label{E:declination3}
    \cos\theta &=& \cos\theta_*\;\cos\left(S_{\theta}S_r \sqrt{\mathcal{C}}\int_{r_*}^{r}\frac{\d\bar{r}}{\bar{r}^2\sqrt{X(\bar{r})}}\right) 
    = \cos\theta_*\;\cos\left(\sqrt{\mathcal{C}}\int_{r_*}^{r}\frac{S_r \, \d\bar{r}}{\bar{r}^2\sqrt{X(\bar{r})}}\right) \ , \nonumber
\end{eqnarray}
with the last step coming from the fact that $\cos(\cdots)$ is an even function of its argument.
That is
\begin{eqnarray}
\label{E:declination4}
    \cos\theta &=& \cos\theta_*\;\cos\left(\sqrt{\mathcal{C}}\int_{r_*}^{r}\frac{|\d\bar{r}|}{\bar{r}^2\sqrt{X(\bar{r})}}\right) \ ,
\end{eqnarray}\enlargethispage{40pt}
where the phase 
\begin{equation}
\hbox{(phase)} = \sqrt{\mathcal{C}}\int_{r_*}^{r}\frac{|\d\bar{r}|}{\bar{r}^2\sqrt{X(\bar{r})}}
\end{equation}
is monotone increasing.

One is again reduced to investigating yet another ultra-elliptic integral~\cite{hyper,hyper2}, with the declination 
angle $\theta$ oscillating periodically as a function of this phase with period $2\pi$, and with each ``bounce'' from $r_{min}$ to $r_{max}$ advancing the phase of the cosine by an amount
\begin{equation}
\Delta \hbox{(phase)}= \sqrt{\mathcal{C}}\int_{r_{min}}^{r_{max}}\frac{\d\bar{r}}{\bar{r}^2\sqrt{X(\bar{r})}}\ .
\end{equation}
As before, this equation cannot be explicitly integrated in closed form.

%==============================
\subsection{Integrating the azimuth}
\label{SS:azimuth}
%==============================

\noindent We now finally consider the ODE for the evolution of the azimuthal angle: $\d\phi/\d\lambda$. (This particular sub-case is considerably messier than the previous ones.) Using equations~(\ref{E:phi}) and (\ref{E:r}) we find

\begin{equation}
    \frac{\d\phi}{\d r} = S_rS_\phi\left[2J\frac{\sqrt{2mr}}{r^3(r-2m)}\right] + S_r\left[\frac{L}{r^2\sin^2\theta\sqrt{X(r)}}+2J\frac{E-2JL/r^3}{r^3(1-2m/r)\sqrt{X(r)}}\right] \ .
\end{equation}
Consequently
\begin{eqnarray}
\label{E:phi(r)_general}
    \phi(r) &=& \phi_0 + S_rS_\phi\left\lbrace 2J\int_{r_0}^{r} \frac{\sqrt{2m\bar{r}}}{\bar{r}^3(\bar{r}-2m)}\,\d\bar{r}\right\rbrace \nonumber \\
    && \nonumber \\
    && \quad + S_r \int_{r_0}^{r}\left(\frac{L}{\bar{r}^2\sin^2[\theta(\bar{r})]\sqrt{X(\bar{r})}} +\frac{2J(E-2JL/\bar{r}^3)}{\bar{r}^3(1-2m/\bar{r})\sqrt{X(\bar{r})}}\right)\,\d\bar{r} \ ,
    \qquad
\end{eqnarray}
where as per our previous discussion we have $S_r=\mbox{sign}(r-r_0)$.
The last term appearing here is again an ultra-elliptic integral~\cite{hyper,hyper2}. In contrast, the penultimate term is somewhat worse, because the integrand  contains $\theta(\bar r)$, the overall integral is an \emph{iteration} of an ultra-elliptic integral. 

We can explicitly integrate the first integral in closed form:
\begin{eqnarray}\label{tanh2}
2J\int_{r_0}^r \frac{\sqrt{2m\bar{r}}}{\bar{r}^3(\bar{r}-2m)}\,\d\bar r &=& 2J\sqrt{\frac{2}{m}}\left[\frac{1}{\sqrt{r}}\left(\frac{1}{3r}+\frac{1}{2m}\right)-\frac{1}{\sqrt{r_0}}\left(\frac{1}{3r_0}+\frac{1}{2m}\right)\right] \nonumber \\
&& \nonumber \\
&& \quad + \frac{J}{2m^2}\ln\left[\frac{(\sqrt{r}-\sqrt{2m})(\sqrt{r_0}+\sqrt{2m})}{(\sqrt{r_0}-\sqrt{2m})(\sqrt{r}+\sqrt{2m})}\right] \ .\qquad
\end{eqnarray}
Thence, in general
\begin{eqnarray}\label{phi(r)3}
    \phi(r) &=& \phi_0 + S_rS_\phi\Bigg\lbrace\frac{J}{2m^2}\ln\left[\frac{(\sqrt{r}-\sqrt{2m})(\sqrt{r_0}+\sqrt{2m})}{(\sqrt{r_0}-\sqrt{2m})(\sqrt{r}+\sqrt{2m})}\right] \nonumber \\
    && \nonumber \\
    && + 2J\sqrt{\frac{2}{m}}\left[\frac{1}{\sqrt{r}}\left(\frac{1}{3r}+\frac{1}{2m}\right)-\frac{1}{\sqrt{r_0}}\left(\frac{1}{3r_0}+\frac{1}{2m}\right)\right]\Bigg\rbrace \nonumber \\
    && \nonumber \\
    && + S_r \int_{r_0}^{r}\left(\frac{L}{\bar{r}^2\sin^2[\theta(\bar{r})]\sqrt{X(\bar{r})}} +\frac{2J(E-2JL/\bar{r}^3)}{\bar{r}^3(1-2m/\bar{r})\sqrt{X(\bar{r})}}\right)\,\d\bar{r} \ . \label{phi(r)}
\end{eqnarray}
As before, this equation being ultra-elliptic means it cannot be explicitly integrated in closed fully analytic form,
at least not in terms of elementary functions.

%=========================================
\subsection{Summary of generic geodesic evolution}
\label{SS:summary-generic-geodesic}
%=========================================

Overall, while these generic geodesic equations cannot be integrated in closed and fully analytic form, they are still integrable in the formal technical sense. 
In terms of complete ultra-elliptic integrals some of the key quantities of interest (both mathematically and physically) are the ``periods'':
\begin{equation}
\int_{r_{min}}^{r_{max}} {\d\bar r\over\sqrt{X(\bar r)}}; \qquad
%\end{equation}
%\begin{equation}
\int_{r_{min}}^{r_{max}}\frac{\d\bar{r}}{\bar{r}^2\sqrt{X(\bar{r})}};
\end{equation}
and
\begin{equation}
\int_{r_{min}}^{r_{max}} \frac{E-2JL/\bar{r}^3}{(1-2m/\bar{r})\sqrt{X(\bar{r})}} \; \d\bar r; 
\qquad
\int_{r_{min}}^{r_{max}} \frac{E-2JL/\bar{r}^3}{\bar r^3 (1-2m/\bar{r})\sqrt{X(\bar{r})}} \; \d\bar r; 
\end{equation}
and the iterated integral
\begin{equation}
\int_{r_{min}}^{r_{max}} \frac{L}{\bar{r}^2\sin^2[\theta(\bar{r})]\sqrt{X(\bar{r})}} \,\d\bar{r} \ . 
\end{equation}
Taylor expanding $(1-2m/r)^{-1}$ focusses attention on the quantities
\begin{equation}
\int_{r_{min}}^{r_{max}} {\d\bar r\over \bar r^n \sqrt{X(\bar r)}}; \qquad (n\in\mathbb{N}).
\end{equation}
Typically, the ``periods'' of these ultra-elliptic integrals will be incommensurate.

If further specific constraints are now imposed, then these equations can indeed be integrated in closed form. In the next section we explicate both the null and timelike geodesics for when $\mathcal{C}=0$ in closed fully analytic form. We are able to recover the ``rain" geodesics from~\cite{4-of-us}, as well as present a simple derivation of both the ``drip" and ``hail" geodesics. 
%
%=====================================================
\section{Geodesics with Carter constant zero}
%=====================================================
\label{S:Carter-zero}
%=====================================================

If the Carter constant is zero then the geodesic equations simplify radically. Firstly, if $\mathcal{C}=0$ then from the manifest positivity of
\begin{equation} 
\mathcal{C}=\left(r^2\frac{\d\theta}{\d\lambda}\right)^2 +\left(\frac{L}{\sin\theta}\right)^2 \ ,
\end{equation}
we see that we must have \emph{both} $L=0$ \emph{and} $\d\theta/\d\lambda\equiv 0$. The condition that $L=0$ physically constrains the geodesics to the trajectories of ZAMOs (zero angular momentum observers), whilst $\d\theta/\d\lambda=0\Longrightarrow\theta(r)=\theta_0$; some constant $\theta_0$. 

Furthermore our expression for $X(r)$ significantly simplifies since now:
\begin{equation}
X(r) = E^2+\epsilon\left( 1-\frac{2m}{r}\right) \ .
\end{equation}

We find that the four trajectory equations reduce to:
\begin{equation}
\frac{\d r}{\d\lambda}= S_r \sqrt{X(r)} \ ;
\label{E:r2}
\end{equation}
\begin{equation}
\frac{\d t}{\d\lambda} = \frac{E+S_r \sqrt{(2m/r)X(r)}}{1-2m/r} \ ;
\label{E:t2}
\end{equation}
\begin{equation}
\frac{\d\theta}{\d\lambda}=0 \ ;
\label{E:theta2}
\end{equation}
%and
\begin{equation}
\frac{\d\phi}{\d\lambda} = \frac{2J}{r^3}\left(\frac{E+S_\phi\sqrt{(2m/r)X(r)}}{1-2m/r}\right) \ .
\label{E:phi2}
\end{equation}
The form of these equations suggests it would be particularly useful to separate our analysis of null geodesics (photons) from timelike geodesics (massive particles).

%=============================
\subsection{Null geodesics (photons) with Carter constant zero}
\label{photonsC0}
%=============================

For null geodesics (photons)  with Carter constant zero, we have the following conditions:
\begin{equation}
\mathcal{C}=0 \ ; \qquad L=0 \ ; \qquad  \theta(r)=\theta_0 \ ; \qquad
X(r) = E^2 \ .
\end{equation}
Furthermore, without loss of generality, we can rescale the affine parameter to set $E\to1$, so that $X(r)\to1$.
From equation \eqref{lambda_r}, we now find
\begin{equation}
\lambda(r) = \lambda_0 + S_r \int_{r_0}^r {\d\bar r\over\sqrt{X(\bar r)}}= 
\lambda_0 + S_r \int_{r_0}^r {\d\bar r}= \lambda_0 + S_r (r-r_0) \ .
\end{equation}
That is, we find the simple linear relation
\begin{equation}
    r(\lambda) = S_r (\lambda-\lambda_0)+r_0 \ .
\end{equation}
Thus in this particular situation $r$ is an affine parameter. Ingoing geodesics will crash into the central singularity in finite affine time, whereas outgoing geodesics can emerge from the horizon ($r=2m$) at finite affine time (which we shall soon see corresponds to minus infinity in Killing time). The apparent asymmetry between ingoing and outgoing null geodesics is a side-effect of the initial choices made in setting up the Painlev\'e--Gullstrand coordinate system.
(Did one choose an ingoing or outgoing Painlev\'e--Gullstrand coordinate system?)

Furthermore, equation (\ref{E:t(r)}) for $t(r)$ now reduces to
\begin{eqnarray}
t(r) = t_0 + 2\sqrt{2m}(\sqrt{r}-\sqrt{r_0}) &+& 2m\ln\left[\frac{(\sqrt{r}-\sqrt{2m})(\sqrt{r_0}+\sqrt{2m})}{(\sqrt{r_0}-\sqrt{2m})(\sqrt{r}+\sqrt{2m})}\right] \nonumber \\
&& \nonumber \\
&& \qquad \qquad \qquad + S_r  \int_{r_{0}}^{r}\frac{\bar{r}}{\bar{r}-2m}\, \d\bar{r} \ .
\end{eqnarray}
This integrates explicitly to
\begin{eqnarray}
    t(r) = t_0 + 2\sqrt{2m}(\sqrt{r}-\sqrt{r_0}) &+& 2m\ln\left[\frac{(\sqrt{r}-\sqrt{2m})(\sqrt{r_0}+\sqrt{2m})}{(\sqrt{r_0}-\sqrt{2m})(\sqrt{r}+\sqrt{2m})}\right] \nonumber \\
    && + S_r\left[r-r_0+2m\ln\left(\frac{r-2m}{r_0-2m}\right)\right] \ .
\end{eqnarray}
This can also be rewritten as
\begin{eqnarray}
    t(r) &=& t_0 + 2\sqrt{2m}(\sqrt{r}-\sqrt{r_0}) + \vert r-r_0\vert \nonumber \\[3pt]
    && +2m\left\lbrace S_r \ln\left[\frac{r+2m-2\,S_r  \sqrt{2mr}}{r_0+2m-2\,S_r  \sqrt{2mr_0}}\right] \right\rbrace \ ,
\end{eqnarray}
or even
\begin{eqnarray}
    t(r) &=& t_0 + 2\sqrt{2m}(\sqrt{r}-\sqrt{r_0}) + \vert r-r_0\vert \nonumber 
     +4m\left\lbrace S_r \ln\left[\frac{\sqrt{r}-S_r  \sqrt{2m}}{\sqrt{r_0}-S_r  \sqrt{2m}}\right] \right\rbrace \ .
\end{eqnarray}
Which form one uses is really a matter of taste. 
\enlargethispage{30pt}

Finally, from equation (\ref{phi(r)3}) we also have
\begin{eqnarray}
\phi(r) &=& \phi_0 + S_rS_\phi\,\Bigg\lbrace \frac{J}{2m^2}\ln\left[\frac{(\sqrt{r}-\sqrt{2m})(\sqrt{r_0}+\sqrt{2m})}{(\sqrt{r_0}-\sqrt{2m})(\sqrt{r}+\sqrt{2m})}\right] \nonumber \\
    && \nonumber \\
    && + 2J\sqrt{\frac{2}{m}}\left[\frac{1}{\sqrt{r}}\left(\frac{1}{3r}+\frac{1}{2m}\right)-\frac{1}{\sqrt{r_0}}\left(\frac{1}{3r_0}+\frac{1}{2m}\right)\right]\Bigg\rbrace \nonumber \\
    && \nonumber \\
    && + 2J\,S_r \int_{r_0}^{r}\frac{\d\bar{r}}{\bar{r}^3(1-2m/\bar{r})} \ .
\end{eqnarray}
This integrates explicitly to
\begin{eqnarray}
\phi(r) &=& \phi_0 + S_rS_\phi\,\Bigg\lbrace \frac{J}{2m^2}\ln\left[\frac{(\sqrt{r}-\sqrt{2m})(\sqrt{r_0}+\sqrt{2m})}{(\sqrt{r_0}-\sqrt{2m})(\sqrt{r}+\sqrt{2m})}\right] \nonumber \\
    && \nonumber \\
    && + 2J\sqrt{\frac{2}{m}}\left[\frac{1}{\sqrt{r}}\left(\frac{1}{3r}+\frac{1}{2m}\right)-\frac{1}{\sqrt{r_0}}\left(\frac{1}{3r_0}+\frac{1}{2m}\right)\right]\Bigg\rbrace \nonumber \\
    && \nonumber \\
    && + S_r \,\frac{J}{m}\left\lbrace \frac{1}{r}-\frac{1}{r_0} + \frac{1}{2m}\ln\left[\frac{(1-2m/r)}{(1-2m/r_0)}\right]\right\rbrace \ .
\end{eqnarray}
Notice that while these equations are rather complicated, they are all fully explicit and given in terms of elementary functions. We also note that these equations have sensible limiting behaviour; as $r\rightarrow r_0$, we have that $\lambda(r),\, t(r),\, \phi(r)\rightarrow \lambda_{0},\, t_{0},\, \phi_{0}$ respectively.\\

\enlargethispage{20pt}
%=============================
\subsection{Timelike geodesics (massive particles) with Carter constant zero}
%=============================

For timelike geodesics (massive particles) with Carter constant zero, we have the following conditions:
\begin{equation}
\label{timelikeconditions}
    \mathcal{C} = 0 \ ; \qquad L=0 \ ; \qquad \theta(r)=\theta_0 \ ; \qquad X(r) = E^2-1+\frac{2m}{r} \ .
\end{equation}
The form of the polynomial $X(r)$ suggests that we should split our analysis into three cases. For the first case we set $E=1$ (since in this case $X(r)$ reduces significantly; $X(r)\to2m/r$), for the second case we set $E>1$, and for the third case we set $E<1$. Physically, geodesics with $E=1$ represent particles that have zero radial velocity at spatial infinity (these are \emph{marginally bound} geodesics). Geodesics with $E>1$ represent particles that have non-zero velocity at spatial infinity (these are \emph{unbound} geodesics). Geodesics with $E<1$ represent particles that are in \emph{bound} orbits, and so never escape to spatial infinity.
For ingoing geodesics, these correspond to the ``rain'' ($E=1$), ``hail'' ($E>1$), and ``drip'' ($E<1$) geodesics.

%=============================
\subsubsection{Marginal geodesics $E=1$}
%=============================

Setting $E=1$, our conditions reduce to:
\begin{equation}
\mathcal{C}=0 \ ; \quad L=0 \ ; \quad \theta(r)=\theta_0 \ ; \quad X(r)=\frac{2m}{r} \ .
\end{equation}
So, for our expression for $\lambda(r)$ we find
\begin{eqnarray} \label{lamda(r)_E_eq_1}
    \lambda(r) = \lambda_0 + S_r \int_{r_0}^{r}\frac{\d\bar{r}}{\sqrt{X(\bar{r})}} &=& \lambda_0 + S_r \int_{r_0}^{r}\frac{\d\bar{r}}{\sqrt{2m/\bar{r}}} \nonumber \\
    && \nonumber \\
    &=& \lambda_0 + S_r \,\sqrt{\frac{2}{m}}\left[\frac{r^{\frac{3}{2}}-r_0^{\frac{3}{2}}}{3}\right] \ .
\end{eqnarray}
Hence
\begin{equation}
    r(\lambda) = \left\lbrace \frac{3\sqrt{2m}}{2}\left[S_r (\lambda-\lambda_0)\right]+r_0^{\frac{3}{2}}\right\rbrace^{\frac{2}{3}} \ .
\end{equation}
Our general expression (\ref{E:t(r)}) for $t(r)$ reduces to
\begin{eqnarray}
    t(r) = t_0 + 2\sqrt{2m}\left(\sqrt{r}-\sqrt{r_0}\,\right) &+& 2m\ln\left[\frac{(\sqrt{r}-\sqrt{2m})(\sqrt{r_0}+\sqrt{2m})}{(\sqrt{r_0}-\sqrt{2m})(\sqrt{r}+\sqrt{2m})}\right] \nonumber \\
    && \nonumber \\
    &+& S_r \int_{r_0}^{r}\frac{\d\bar{r}}{(1-2m/\bar{r})\sqrt{2m/\bar{r}}} \ .
\end{eqnarray}
We may now compute the somewhat unwieldy integral
\begin{eqnarray}
    \int_{r_0}^{r}\frac{\d\bar{r}}{(1-2m/\bar{r})\sqrt{2m/\bar{r}}} &=& 2\sqrt{2m}(\sqrt{r}-\sqrt{r_0}) + \sqrt{\frac{2}{m}}\left(\frac{r^{\frac{3}{2}}-r_0^{\frac{3}{2}}}{3}\right) \nonumber \\
    && \nonumber \\
    && + 2m\ln\left[\frac{(\sqrt{r}-\sqrt{2m})(\sqrt{r_0}+\sqrt{2m})}{(\sqrt{r_0}-\sqrt{2m})(\sqrt{r}+\sqrt{2m})}\right] \ ,
\end{eqnarray}
giving the following explicit form for $t(r)$
\begin{eqnarray}\label{t(r)E1}
    t(r) = t_0 &+& 2\sqrt{2m}(\sqrt{r}-\sqrt{r_0}) + 2m\ln\left[\frac{(\sqrt{r}-\sqrt{2m})(\sqrt{r_0}+\sqrt{2m})}{(\sqrt{r_0}-\sqrt{2m})(\sqrt{r}+\sqrt{2m})}\right] \nonumber \\
    && \nonumber \\
    && \qquad + S_r\Bigg\lbrace 2\sqrt{2m}(\sqrt{r}-\sqrt{r_0}) + \sqrt{\frac{2}{m}}\left(\frac{r^{\frac{3}{2}}-r_0^{\frac{3}{2}}}{3}\right) \nonumber \\
    && \nonumber \\
    && \qquad \qquad \qquad + 2m\ln\left[\frac{(\sqrt{r}-\sqrt{2m})(\sqrt{r_0}+\sqrt{2m})}{(\sqrt{r_0}-\sqrt{2m})(\sqrt{r}+\sqrt{2m})}\right]\Bigg\rbrace \ .
\end{eqnarray}
It is worth noting that for $S_r =-1$ (corresponding to ingoing geodesics) we find the particularly simple result
\begin{eqnarray}
\label{E1_t(r)_1}
    t(r) = t_0 - \sqrt{\frac{2}{m}}\left(\frac{r^{\frac{3}{2}}-r_0^{\frac{3}{2}}}{3}\right) \ ,
\end{eqnarray}
whilst for $S_r =+1$ (corresponding to outgoing geodesics) we obtain
\begin{eqnarray}
\label{E1_t(r)_2}
%\begin{split}
    t(r) &=& t_0  +\sqrt{\frac{2}{m}}\left(\frac{r^{\frac{3}{2}}-r_0^{\frac{3}{2}}}{3}\right) 
    + 4\sqrt{2m}\left(\sqrt{r}-\sqrt{r_0}\right) 
    \nonumber\\
    && + 4m\ln\left[\frac{(\sqrt{r}-\sqrt{2m})(\sqrt{r_0}+\sqrt{2m})}{(\sqrt{r_0}-\sqrt{2m})(\sqrt{r}+\sqrt{2m})}\right] \ .
%\end{split}
\end{eqnarray}
The apparent asymmetry between ingoing and outgoing timelike geodesics is a side-effect of the choices made in setting up the Painlev\'e--Gullstrand coordinate system. Note that in this situation the formulae for $t(r)$ are $J$-independent, so this apparent asymmetry is already present for Schwarzschild spacetime in  Painlev\'e--Gullstrand coordinates. The usual choices allow ingoing geodesics to penetrate the horizon and crash into the central singularity in finite Killing time, while outgoing geodesics need an infinite amount of Killing time to escape from the horizon at $r=2m$ and so get `stuck'.

Our general expression (\ref{phi(r)3}) for $\phi(r)$ reduces to
\begin{eqnarray}
    \phi(r) &=& \phi_0 + S_rS_\phi\,\Bigg\lbrace \frac{J}{2m^2}\ln\left[\frac{(\sqrt{r}-\sqrt{2m})(\sqrt{r_0}+\sqrt{2m})}{(\sqrt{r_0}-\sqrt{2m})(\sqrt{r}+\sqrt{2m})}\right] \nonumber \\
    && \nonumber \\
    && + 2J\sqrt{\frac{2}{m}}\left[\frac{1}{\sqrt{r}}\left(\frac{1}{3r}+\frac{1}{2m}\right)-\frac{1}{\sqrt{r_0}}\left(\frac{1}{3r_0}+\frac{1}{2m}\right)\right]\Bigg\rbrace \nonumber \\
    && \nonumber \\
    && + 2J\,S_r \int_{r_0}^{r}\frac{\d\bar{r}}{\bar{r}^3(1-2m/\bar{r})\sqrt{2m/\bar{r}}} \ .
\end{eqnarray}
We may integrate the remaining integral in closed form
\begin{eqnarray}
    \int_{r_0}^{r}\frac{\d\bar{r}}{\bar{r}^3(1-2m/\bar{r})\sqrt{2m/\bar{r}}} &=& \frac{1}{m\sqrt{2m}}\left(\frac{1}{\sqrt{r}}-\frac{1}{\sqrt{r_0}}\right) \nonumber \\
    && \nonumber \\
    && + \frac{1}{4m^2}\ln\left[\frac{(\sqrt{r}-\sqrt{2m})(\sqrt{r_0}+\sqrt{2m})}{(\sqrt{r_0}-\sqrt{2m})(\sqrt{r}+\sqrt{2m})}\right] \ .
\end{eqnarray}
This gives the fully explicit form for $\phi(r)$ as:
\begin{eqnarray}\label{phi(r)timelikeE=1}
    \phi(r) &=& \phi_0 + S_rS_\phi\,\Bigg\lbrace \frac{J}{2m^2}\ln\left[\frac{(\sqrt{r}-\sqrt{2m})(\sqrt{r_0}+\sqrt{2m})}{(\sqrt{r_0}-\sqrt{2m})(\sqrt{r}+\sqrt{2m})}\right] \nonumber \\
    && \nonumber \\
    && + 2J\sqrt{\frac{2}{m}}\left[\frac{1}{\sqrt{r}}\left(\frac{1}{3r}+\frac{1}{2m}\right)-\frac{1}{\sqrt{r_0}}\left(\frac{1}{3r_0}+\frac{1}{2m}\right)\right]\Bigg\rbrace \nonumber \\
    && \nonumber \\
    && + S_r \left\lbrace \frac{2J}{m\sqrt{2m}}\left(\frac{1}{\sqrt{r}}-\frac{1}{\sqrt{r_0}}\right) + \frac{J}{2m^2}\ln\left[\frac{(\sqrt{r}-\sqrt{2m})(\sqrt{r_0}+\sqrt{2m})}{(\sqrt{r_0}-\sqrt{2m})(\sqrt{r}+\sqrt{2m})}\right]\right\rbrace \ . \nonumber \\
    &&
\end{eqnarray}
It is now straightforward to recover the ``rain" geodesics explored in~\cite{4-of-us}, which model a ZAMO dropped from spatial infinity with zero initial radial velocity. This physical scenario requires $\C=0$, $L=0$, $\epsilon=-1$, and $E=1$, as above, and furthermore is an \emph{ingoing retrograde} geodesic (corresponding mathematically to fixing $S_r=S_\phi=-1$). These geodesics are retrograde since we must give them an initial nonzero angular velocity in the retrograde direction at spatial infinity in order for $L=0$ to hold along the length of the geodesic. From equations \eqref{E1_t(r)_1} and \eqref{phi(r)timelikeE=1}, we find the explicit ``rain" geodesics:
\begin{equation}\label{t(r)_E=1}
t(r) = t_0 - \sqrt{\frac{2}{m}}\left(\frac{r^{\frac{3}{2}}-r_0^{\frac{3}{2}}}{3}\right) \ ,
\end{equation}
which we may re-phrase (defining $t_{crash}$ to be the amount of elapsed Killing time for an ingoing geodesic to crash into the central singularity) as
\begin{equation}
   t_{rain}(r) = t_{crash} - \sqrt{\frac{2}{m}}\;  \frac{r^{\frac{3}{2}}} {3} \ .
\end{equation}
Similarly
\begin{equation}\label{phi(r)_E=1}
\phi(r) = \phi_0 + \frac{2J}{3}\sqrt{\frac{2}{m}}\left(\frac{1}{r^{\frac{3}{2}}}-\frac{1}{r_0^{\frac{3}{2}}}\right) \ ,
\end{equation}
which we may rephrase as
\begin{equation}\label{phi(r)_E=1b}
\phi_{rain}(r) = \phi_\infty + \frac{2J}{3}\sqrt{\frac{2}{m}}\;\;
\frac{1}{r^{\frac{3}{2}}}\ .
\end{equation}
If we differentiate with respect to $r$, we find
\begin{equation}
\frac{\d t}{\d r} = -\sqrt{\frac{r}{2m}} \ ,
\qquad\hbox{and}\qquad
\frac{\d\phi}{\d r} = -\frac{2J}{\sqrt{2m}}\, r^{-5/2} \ , 
\end{equation}
which are exactly the equations defining the rain geodesics as given in reference~\cite{4-of-us}.
%
%=============================
\subsubsection{Unbound geodesics $E>1$}
%=============================

For $E>1$, let us first restore the full list of conditions from equation (\ref{timelikeconditions}):
\begin{equation}
\mathcal{C}=0 \ ; \quad L=0 \ ; \quad \theta(r)=\theta_0 \ ; \quad X(r)=E^2-1+\frac{2m}{r} \ .
\end{equation}
Notice that for $E>1$, $r$ is \emph{a priori} unconstrained; $r\in(0,+\infty)$.
\enlargethispage{20pt}

From our  general result (\ref{lambda_r}) for $\lambda(r)$ we have
\begin{equation}
\begin{split}
\lambda(r) & = \lambda_0+S_r \int_{r_0}^r \frac{\d\Bar{r}}{\sqrt{X(\Bar{r})}} = \lambda_0+S_r \int_{r_0}^r \frac{\d\Bar{r}}{\sqrt{E^2-1+2m/\Bar{r}}} \\
& \\
& = \lambda_0 + \frac{S_r }{E^2-1}\Bigg\lbrace r\sqrt{E^2-1+2m/r}-r_0\sqrt{E^2-1+2m/r_0} \\
& \hspace{1cm} +\frac{m}{\sqrt{E^2-1}}\left[\ln{r_0\over r} + 2 \ln\left|
\frac{\sqrt{E^2-1} +\sqrt{E^2-1+2m/r_0}}
{\sqrt{E^2-1} +\sqrt{E^2-1+2m/r}}
\right|\right]\Bigg\rbrace \ .
\end{split}
\end{equation}
\enlargethispage{20pt}
Conducting a Taylor series expansion around $E=1$, we find
\begin{equation}
\lambda(r) = \lambda_0 + S_r \sqrt{\frac{2}{m}}\left(\frac{r^{3/2}-r_0^{3/2}}{3}\right) + \O(E-1) \ .
\end{equation}
In the limit where $E\rightarrow 1$, we see that our expression for $\lambda(r)$ simplifies to \eqref{lamda(r)_E_eq_1}, as expected.

The unwieldy integral in our general expression (\ref{tanh1}) for $t(r)$ now becomes
\begin{equation}
\begin{split}
\int_{r_0}^r & \frac{E}{(1-2m/\bar{r})\sqrt{E^2-1+2m/\bar{r}}}\,\d\Bar{r} \\
& =\frac{E}{E^2-1}\left(r\sqrt{E^2-1+2m/r}-r_0\sqrt{E^2-1+2m/r_0}\right) \\
& \hspace{0.4cm} -2m\ln\left[\frac{(2m-r_0)\left(r-2m-2Er\left(E+\sqrt{E^2-1+2m/r}\right)\right)}{(2m-r)\left(r_0-2m-2Er_0\left(E+\sqrt{E^2-1+2m/r_0}\right)\right)}\right] \\
& \hspace{0.4cm} +\frac{Em(2E^2-3)}{(E^2-1)^{3/2}}\ln\left[\frac{m-r+E^2r+r\sqrt{(E^2-1)(E^2-1+2m/r)}}{m-r_0+E^2r_0+r_0\sqrt{(E^2-1)(E^2-1+2m/r_0)}}\right] \ .
\end{split}
\end{equation}
This yields the following fully explicit result for $t(r)$
\begin{eqnarray}
\label{E:t(r)_C=0_E>1}
    t(r) &=& t_0 + 2\sqrt{2m}\left(\sqrt{r}-\sqrt{r_0}\right) + 2m\ln\left[\frac{(\sqrt{r}-\sqrt{2m})(\sqrt{r_0}+\sqrt{2m})}{(\sqrt{r_0}-\sqrt{2m})(\sqrt{r}+\sqrt{2m})}\right] \nonumber \\
    && \nonumber \\
    && +S_r \Bigg\lbrace \frac{E}{E^2-1}\left(r\sqrt{E^2-1+2m/r}-r_0\sqrt{E^2-1+2m/r_0}\right) \nonumber \\
    && \nonumber \\
    && \hspace{0.6cm} -2m\ln\left[\frac{(2m-r_0)\left(r-2m-2Er\left(E+\sqrt{E^2-1+2m/r}\right)\right)}{(2m-r)\left(r_0-2m-2Er_0\left(E+\sqrt{E^2-1+2m/r_0}\right)\right)}\right]\nonumber \\
    && \nonumber \\
    && \hspace{0.6cm} +\frac{Em(2E^2-3)}{(E^2-1)^{3/2}}\ln\left[\frac{m-r+E^2r+r\sqrt{(E^2-1)(E^2-1+2m/r)}}{m-r_0+E^2r_0+r_0\sqrt{(E^2-1)(E^2-1+2m/r_0)}}\right]\Bigg\rbrace \ . \nonumber \\
    &&
\end{eqnarray}
Conducting a Taylor series expansion around $E=1$ gives
\begin{eqnarray}
    t(r) = t_0 &+& 2\sqrt{2m}(\sqrt{r}-\sqrt{r_0}) + 2m\ln\left[\frac{(\sqrt{r}-\sqrt{2m})(\sqrt{r_0}+\sqrt{2m})}{(\sqrt{r_0}-\sqrt{2m})(\sqrt{r}+\sqrt{2m})}\right] \nonumber \\
    && \nonumber \\
    && \qquad + S_r \Bigg\lbrace 2\sqrt{2m}(\sqrt{r}-\sqrt{r_0}) + \sqrt{\frac{2}{m}}\left(\frac{r^{\frac{3}{2}}-r_0^{\frac{3}{2}}}{3}\right) \nonumber \\
    && \nonumber \\
    && \qquad \qquad \qquad + 2m\ln\left[\frac{(\sqrt{r}-\sqrt{2m})(\sqrt{r_0}+\sqrt{2m})}{(\sqrt{r_0}-\sqrt{2m})(\sqrt{r}+\sqrt{2m})}\right]\Bigg\rbrace + \mathcal{O}(E-1) \ . \nonumber \\
    &&
\end{eqnarray}
Hence in the limit $E\rightarrow 1$, we have 
\begin{eqnarray}
    t(r) = t_0 &+& 2\sqrt{2m}(\sqrt{r}-\sqrt{r_0}) + 2m\ln\left[\frac{(\sqrt{r}-\sqrt{2m})(\sqrt{r_0}+\sqrt{2m})}{(\sqrt{r_0}-\sqrt{2m})(\sqrt{r}+\sqrt{2m})}\right] \nonumber \\
    && \nonumber \\
    && \qquad + S_r \Bigg\lbrace 2\sqrt{2m}(\sqrt{r}-\sqrt{r_0}) + \sqrt{\frac{2}{m}}\left(\frac{r^{\frac{3}{2}}-r_0^{\frac{3}{2}}}{3}\right) \nonumber \\
    && \nonumber \\
    && \qquad \qquad \qquad + 2m\ln\left[\frac{(\sqrt{r}-\sqrt{2m})(\sqrt{r_0}+\sqrt{2m})}{(\sqrt{r_0}-\sqrt{2m})(\sqrt{r}+\sqrt{2m})}\right]\Bigg\rbrace \ ,
\end{eqnarray}
which is just equation \eqref{t(r)E1}, as expected.

%\clearpage
Lastly, the somewhat unwieldy integral in our general expression (\ref{E:phi(r)_general}) for $\phi(r)$ is now given by
\begin{eqnarray}
    && \int_{r_0}^{r}\frac{2EJ}{\bar{r}^3(1-2m/\bar{r})\sqrt{E^2-1+2m/\bar{r}}}\,\d\bar{r} \nonumber \\
    && \nonumber \\
    && \quad = \frac{EJ}{m^2}\left(\sqrt{E^2-1+2m/r}-\sqrt{E^2-1+2m/r_0}\right) \nonumber \\
    && \nonumber \\
    && - \frac{J}{2m^2}\left\lbrace\ln\left[\frac{1-2m/r_0}{1-2m/r}\right]+\ln\left[\frac{2E(E+\sqrt{E^2-1+2m/r})-(1-2m/r)}{2E(E+\sqrt{E^2-1+2m/r_0})-(1-2m/r_0)}\right]\right\rbrace \ , \nonumber \\
    &&
\end{eqnarray}
so we find that $\phi(r)$ reduces to
\begin{eqnarray}\label{B41}
    \phi(r) &=& \phi_0 - S_rS_\phi\,\Bigg\lbrace \frac{J}{2m^2}\ln\left[\frac{(\sqrt{r}-\sqrt{2m})(\sqrt{r_0}+\sqrt{2m})}{(\sqrt{r_0}-\sqrt{2m})(\sqrt{r}+\sqrt{2m})}\right] \nonumber \\
    && \nonumber \\
    && + 2J\sqrt{\frac{2}{m}}\left[\frac{1}{\sqrt{r}}\left(\frac{1}{3r}+\frac{1}{2m}\right)-\frac{1}{\sqrt{r_0}}\left(\frac{1}{3r_0}+\frac{1}{2m}\right)\right]\Bigg\rbrace \nonumber \\
    && \nonumber \\
    && + S_r\,\Bigg\lbrace \frac{EJ}{m^2}\left(\sqrt{E^2-1+2m/r}-\sqrt{E^2-1+2m/r_0}\right) \nonumber \\
    && \nonumber \\
    && - \frac{J}{2m^2}\left[\ln\left(\frac{1-2m/r_0}{1-2m/r}\right)+\ln\left(\frac{2E(E+\sqrt{E^2-1+2m/r})-(1-2m/r)}{2E(E+\sqrt{E^2-1+2m/r_0})-(1-2m/r_0)}\right)\right]\Bigg\rbrace \ . \nonumber \\
    &&
\end{eqnarray}
If we now make the direct substitution $E=1$, we find
\begin{eqnarray}
    \phi(r) &=& \phi_0 - S_rS_\phi\,\Bigg\lbrace \frac{J}{2m^2}\ln\left[\frac{(\sqrt{r}-\sqrt{2m})(\sqrt{r_0}+\sqrt{2m})}{(\sqrt{r_0}-\sqrt{2m})(\sqrt{r}+\sqrt{2m})}\right] \nonumber \\
    && \nonumber \\
    && + 2J\sqrt{\frac{2}{m}}\left[\frac{1}{\sqrt{r}}\left(\frac{1}{3r}+\frac{1}{2m}\right)-\frac{1}{\sqrt{r_0}}\left(\frac{1}{3r_0}+\frac{1}{2m}\right)\right]\Bigg\rbrace \nonumber \\
    && \nonumber \\
    && + S_r\,\left\lbrace \frac{2J}{m\sqrt{2m}}\left(\frac{1}{\sqrt{r}}-\frac{1}{\sqrt{r_0}}\right) + \frac{J}{2m^2}\ln\left[\frac{(\sqrt{r}-\sqrt{2m})(\sqrt{r_0}+\sqrt{2m})}{(\sqrt{r_0}-\sqrt{2m})(\sqrt{r}+\sqrt{2m})}\right]\right\rbrace \ , \nonumber \\
    &&
\end{eqnarray}
which is just equation \eqref{phi(r)timelikeE=1}, as expected.

\enlargethispage{20pt}
Overall, while these equations of motion are rather long, they are fully explicit and have appropriate limits when $E\rightarrow 1$.

In all generality, geodesics with $E>1$ are unbound geodesics. When (as herein) the Carter constant is zero, the ingoing  geodesics are the so-called ``hail" geodesics, modelling a ZAMO fired in from spatial infinity with a nonzero initial velocity. These geodesics are ingoing retrograde, and are hence given by
\begin{equation}
    \frac{\d t}{\d r} = \frac{\sqrt{2mr}}{r-2m} - \,\frac{E}{(1-2m/r)\sqrt{E^2-1+2m/r}} \ ,
\end{equation}
and
\begin{equation}
    \frac{\d\phi}{\d r} = - \,\frac{2J\sqrt{2mr}}{r^3(r-2m)} - \,\frac{2EJ}{r^3(1-2m/r)\sqrt{E^2-1+2m/r}} \ .
\end{equation}
%

%=============================================================
\subsubsection{Bound geodesics $E<1$}
%=============================================================

Once again we repeat the full list of conditions from equation~(\ref{timelikeconditions}):

\begin{equation}
    \mathcal{C}=0 \ ; \quad L=0 \ ; \quad \theta(r)=\theta_0 \ ; \quad X(r)=E^2-1+\frac{2m}{r} \ .
\end{equation}
If we let $E<1$, then $X(r)=0$ has a unique root at $r={2m\over1-E^2}$, and in order to keep $\sqrt{X(r)}$ \emph{real}, we see that we have the constraint that $r\in(0,{2m\over1-E^2}]$. In particular $r\leq 2m/(1-E^2)$. Physically, geodesics with $E<1$ are gravitationally bound (and, because $\C=L=0$, must eventually crash into the central singularity). 
These particular bound geodesics are the so-called ``drip" geodesics, corresponding to a ZAMO being dropped from some finite $r_*$ with zero initial velocity. 

For $\lambda(r)$ we have
\begin{equation}
    \lambda(r) = \lambda_0 +S_r\,\int_{r_0}^{r}\frac{\d\bar{r}}{\sqrt{E^2-1+2m/\bar{r}}}\ .
\end{equation}
Explicitly, a brief computation yields
\begin{eqnarray}
    \lambda(r) &= &\lambda_0 +S_r
    \left\{
    {r\sqrt{E^2-1+2m/r}\over 1-E^2} + 
    {m \sin^{-1} \left(1+{E^2r\over m} - {r\over m}\right)
    \over (1-E^2)^{3/2}}
    \right\}
    \nonumber\\
    &&\quad - 
    S_r
    \left\{
    {r_0\sqrt{E^2-1+2m/r_0}\over 1-E^2} + 
    {m \sin^{-1} \left(1+{E^2r_0\over m} - {r_0\over m}\right)
    \over (1-E^2)^{3/2}}
    \right\}\ .
    \end{eqnarray}
At intermediate stages of the computation it is useful to use the identity
 \begin{equation}
\ln(x+iy) = {1\over2}\ln(x^2+y^2) + i \cos^{-1}\left( x\over\sqrt{x^2+y^2}\right).
\end{equation}

For $S_r=+1$ these drip geodesics will crash into the central singularity $r=0$ in finite affine time
\begin{equation}
    \lambda_{crash} = \lambda_0 
    + {m\pi\over2 (1-E^2)^{3/2}} -
    \left\{
    { r_0\sqrt{E^2-1+2m/r_0}\over 1-E^2} + 
    {m \sin^{-1} \left(1+{E^2r_0\over m} - {r_0\over m}\right)
    \over (1-E^2)^{3/2}}
    \right\}
    \ .
    \end{equation}

These ``drip'' geodesics are \emph{qualitatively} (not quantitatively)  somewhat similar to the ``rain" geodesics given by equations~\eqref{t(r)_E=1} and~\eqref{phi(r)_E=1}.

When it comes to evaluating $t(r)$ the result (\ref{E:t(r)_C=0_E>1}) still \emph{formally} holds, but with the understanding that for $E<1$ the trailing term in (\ref{E:t(r)_C=0_E>1}) becomes
\begin{eqnarray}
&&
\frac{1}{(E^2-1)^{3/2}}\ln\left[\frac{m-r+E^2r+r\sqrt{(E^2-1)(E^2-1+2m/r)}}{m-r_0+E^2r_0+r_0\sqrt{(E^2-1)(E^2-1+2m/r_0)}}\right]
\nonumber\\
&& =
\frac{1}{i^3 (1-E^2)^{3/2}}\ln\left[\frac{m-r+E^2r+ir\sqrt{(1-E^2)(E^2-1+2m/r)}}{m-r_0+E^2r_0+i r_0\sqrt{(1-E^2)(E^2-1+2m/r_0)}}\right]
\nonumber\\
&& =
-\frac{1}{ (1-E^2)^{3/2}}\; 
\Bigg\{
\sin^{-1}\left({r\over m}\sqrt{(1-E^2)(E^2-1+2m/r)}\right)
\nonumber\\
&&
\qquad\quad\qquad\qquad\qquad
-
\sin^{-1}\left({r_0\over m}\sqrt{(1-E^2)(E^2-1+2m/r_0)}\right)
\Bigg\}.
\end{eqnarray}
At intermediate stages of the computation it is now useful to use the slightly different identity
 \begin{equation}
\ln(x+iy) = {1\over2}\ln(x^2+y^2) + i \sin^{-1}\left( y\over\sqrt{x^2+y^2}\right).
\end{equation}

This now enforces \emph{manifest reality} of $t(r)$ for $E<1$ and in all its glory we have
\begin{eqnarray}
\label{E:t(r)_C=0_E<1}
    t(r) &=& t_0 + 2\sqrt{2m}\left(\sqrt{r}-\sqrt{r_0}\right) + 2m\ln\left[\frac{(\sqrt{r}-\sqrt{2m})(\sqrt{r_0}+\sqrt{2m})}{(\sqrt{r_0}-\sqrt{2m})(\sqrt{r}+\sqrt{2m})}\right] \nonumber \\
    && \nonumber \\
    && +S_r \Bigg\lbrace \frac{E}{E^2-1}\left(r\sqrt{E^2-1+2m/r}-r_0\sqrt{E^2-1+2m/r_0}\right) \nonumber \\
    && \nonumber \\
    && \hspace{0.6cm} -2m\ln\left[\frac{(2m-r_0)\left(r-2m-2Er\left(E+\sqrt{E^2-1+2m/r}\right)\right)}{(2m-r)\left(r_0-2m-2Er_0\left(E+\sqrt{E^2-1+2m/r_0}\right)\right)}\right]\nonumber \\
    && \nonumber \\
    && \hspace{0.6cm} 
    +\frac{E m (3-2E^2)}{ (1-E^2)^{3/2}}\; 
\Bigg[\sin^{-1}\left({r\over m}\sqrt{(1-E^2)(E^2-1+2m/r)}\right)
\nonumber\\
&& \qquad\qquad\qquad\qquad\qquad 
-
\sin^{-1}\left({r_0\over m}\sqrt{(1-E^2)(E^2-1+2m/r_0)}\right)\Bigg]\Bigg\rbrace \ . \nonumber \\
    &&
\end{eqnarray}

%\enlargethispage{40pt}
Ingoing geodesics $S_r=-1$ (in this context the ``drip'' geodesics) will crash into the central singularity in finite Killing time
\begin{eqnarray}
\label{E:t_crash_C=0_E<1}
    t_{crash} &=& t_0 + 2\sqrt{2mr_0}
    + 2m\ln\left[\frac{(\sqrt{r_0}+\sqrt{2m})}{(\sqrt{r_0}-\sqrt{2m})}\right] 
%    \nonumber \\
%    && \nonumber \\
%    && 
    + \Bigg\lbrace \frac{-Er_0}{1-E^2}\left(\sqrt{E^2-1+2m/r_0}\right) \nonumber \\
    && \nonumber \\
    && \hspace{0.6cm} +2m\ln\left[\frac{(r_0-2m)}{\left(r_0-2m-2Er_0\left(E+\sqrt{E^2-1+2m/r_0}\right)\right)}\right]\Bigg\rbrace \ . 
\end{eqnarray}
There are of course many other ways of rearranging this result.

Finally for $\phi(r)$ the result (\ref{B41})  can be extended to the region $E<1$ without alteration.\enlargethispage{35pt}

%----------------------------------------------------------------------%----------------------------------------------------------------------%========================================
%========================================
%----------------------------------------------------------------------
%----------------------------------------------------------------------
%========================================
\section{Conclusions}
\label{S:conclusions}
%========================================

From this discussion we have seen that, once given the non-trivial Killing tensor for the Lense--Thirring spacetime, we can extract the Carter constant; the fourth constant of the motion. Then the geodesic equations become integrable at least in principle (in terms of ultra-elliptic integrals). This allows us to formally solve for myriads of general geodesics. However, we saw that in full generality, we could not explicitly integrate the equations of motion in closed form, the ultra-elliptic integrals are not ``elementary''. Only when imposing further conditions, such as Carter constant zero, could we then explicitly integrate the equations of motion in algebraically closed form. 

The explicit geodesics given in this discussion are quite tractable and can be applied to a number of astrophysically interesting cases. For example, the Carter constant zero geodesics with $E<1$ are the ``drip" geodesics of the spacetime, while the Carter constant zero geodesics with $E= 1$ are the ``rain'' geodesics, and the Carter constant zero geodesics with $E> 1$ are the
 ``hail" geodesics of the spacetime.

%========================================
\section*{Acknowledgements}
%========================================

JB was supported by a MSc scholarship funded by the Marsden Fund, 
via a grant administered by the Royal Society of New Zealand.
\\
TB was supported by a Victoria University of Wellington MSc scholarship, 
and was also indirectly supported by the Marsden Fund, 
via a grant administered by the Royal Society of New Zealand.
\\
AS was supported by a Victoria University of Wellington PhD Doctoral Scholarship,
and was also indirectly supported by the Marsden fund, 
via a grant administered by the Royal Society of New Zealand.
\\
MV was directly supported by the Marsden Fund, \emph{via} a grant administered by the Royal Society of New Zealand.

%=========================================
%=========================================
%=========================================
%=========================================

%=====================================

\begin{thebibliography}{99}
%=========================================
\newcommand{\arXiv}[1]{arXiv:\href{https://arxiv.org/abs/#1}{\color{blue}#1}}
%This allows using \arXiv{2006.07125} or \arXiv{gr-qc/0009013} for nice links.
%========================================

\bibitem{4-of-us}
Joshua Baines, Thomas Berry, Alex Simpson, and Matt Visser, \\
``Painleve-Gullstrand form of the Lense-Thirring spacetime'',
\\
Universe \textbf{7 \# 4} (2021) 105
\doi{10.3390/universe704010}
[\arXiv{2006.14258} [gr-qc]].
%6 citations counted in INSPIRE as of 29 May 2021
%%%%%%


\bibitem{PGLT2}
J.~Baines, T.~Berry, A.~Simpson and M.~Visser,
``Killing tensor and Carter constant for Painlev\'e-Gullstrand form of Lense-Thirring spacetime'', \\
Universe {\bf 7 \#12}  (2021) 473; 
\doi{10.3390/universe7120473}
[\arXiv{2110.01814} [gr-qc]].
%3 citations counted in INSPIRE as of 03 Dec 2021

%===================================== 
 \bibitem{Unit-lapse}
J.~Baines, T.~Berry, A.~Simpson and M.~Visser,
``Unit-lapse versions of the Kerr spacetime'',
Class. Quant. Grav. \textbf{38} (2021) no.5, 055001
\doi{10.1088/1361-6382/abd071}
[\arXiv{2008.03817} [gr-qc]].
%6 citations counted in INSPIRE as of 05 Jun 2021

%===================================== 
 
\bibitem{painleve1}
Paul Painlev\'e, 
``La m\'ecanique classique et la th\'eorie de la relativit\'e\,", \\
C.~R.~Acad. Sci. (Paris) 173, 677--680(1921).

\bibitem{painleve2}
Paul Painlev\'e, \\
 ``La gravitation dans la m\'ecanique de Newton et dans la m\'ecanique d'Einstein'', \\
 C.~R.~Acad. Sci. (Paris) 173, 873--886(1921).
%\clearpage

\bibitem{gullstrand}
Allvar Gullstrand,\\
``Allgemeine L\"osung des statischen Eink\"orperproblems in der Einsteinschen Gravitationstheorie",\\ 
 Arkiv f\"or Matematik, Astronomi och Fysik. 16 (8): 1--15 (1922).
 
 
 \bibitem{poisson}
K.~Martel and E.~Poisson,\\
``Regular coordinate systems for Schwarzschild and other spherical space-times'',\\
Am. J. Phys. \textbf{69} (2001), 476-480
\doi{10.1119/1.1336836}\\{}
[\arXiv{gr-qc/0001069} [gr-qc]].
%125 citations counted in INSPIRE as of 17 Jun 2020


%-------------------------------------------------------------------

\bibitem{Rajan:2016}
D.~Rajan and M.~Visser,
``Global properties of physically interesting Lorentzian spacetimes'',
Int. J. Mod. Phys. D \textbf{25} (2016) no.14, 1650106
\doi{10.1142/S0218271816501066}
[\arXiv{1601.03355} [gr-qc]].
%4 citations counted in INSPIRE as of 01 Nov 2021

\bibitem{Skakala:2010}
J.~Skakala and M.~Visser,
``The causal structure of spacetime is a parameterized Randers geometry'',
Class. Quant. Grav. \textbf{28} (2011), 065007
\doi{10.1088/0264-9381/28/6/065007}
[\arXiv{1012.4467} [gr-qc]].
%4 citations counted in INSPIRE as of 01 Nov 2021


%=====================================

\bibitem{Birkhoff}
Garret Birkhoff,  % G. D. ,
 \emph{Relativity and Modern Physics}, 
(Harvard University Press, Cambridge, 1923).
% LCCN 23008297. 

\bibitem{Jebsen}
J\o{}rg Tofte Jebsen, ``\"Uber die allgemeinen kugelsymmetrischen 
L\"osungen der Einsteinschen Gravitationsgleichungen im Vakuum'', 
Ark. Mat. Ast. Fys. (Stockholm) {\bf 15} (1921) nr.18. 

\bibitem{Deser}
  Stanley Deser and Joel Franklin,
  ``Schwarzschild and Birkhoff \emph{a la} Weyl'',\\
  Am.\ J.\ Phys.\  {\bf 73} (2005) 261
  [\arXiv{gr-qc/0408067} [gr-qc]].
  %%CITATION = AJPIA,73,261;%%
  
 \bibitem{Ravndal} 
   Nils Voje Johansen, Finn Ravndal, 
   ``On the discovery of Birkhoff's theorem'', 
   Gen.Rel.Grav. 38 (2006) 537-540  
   [\arXiv{physics/0508163} [physics.hist-ph]].
   
\bibitem{Martinez:2004}
E.~Ayon-Beato, C.~Martinez and J.~Zanelli,
``Birkhoff's theorem for three-dimensional AdS gravity'',
Phys. Rev. D \textbf{70} (2004), 044027
\doi{10.1103/PhysRevD.70.044027}
[\arXiv{hep-th/0403227} [hep-th]].
%34 citations counted in INSPIRE as of 03 Dec 2021   
 
\bibitem{Skakala}
J.~Skakala and M.~Visser,\\
``Birkhoff-like theorem for rotating stars in (2+1) dimensions'',\\{}
[\arXiv{0903.2128} [gr-qc]].
%4 citations counted in INSPIRE as of 24 Jun 2020

%======================================
%\clearpage
\bibitem{Lense-Thirring}
Hans Thirring and Josef Lense,  ``\"Uber den Einfluss der Eigenrotation der Zentralk\"orperauf die Bewegung 
der Planeten und Monde nach der Einsteinschen Gravitationstheorie'', Physikalische Zeitschrift, Leipzig Jg. {\bf 19}  (1918), No. 8, p. 156--163.
(Translated in~\cite{Lense-Thirring-translation}.)
%[204--205].\\

\bibitem{Lense-Thirring-translation}
Bahram Mashoon, Friedrich W. Hehl, and Dietmar S. Theiss,
(translators), \\
``On the influence of the proper rotations of central bodies on the motions of planets and moons in Einstein's theory of gravity'', \\
General Relativity and Gravitation  {\bf 16 \# 8} (1984) 727--741.
(Translation of~\cite{Lense-Thirring}.)
%[711].

\bibitem{Pfister}
Herbert Pfister, ``On the history of the so-called Lense--Thirring effect'',
\url{http://philsci-archive.pitt.edu/archive/00002681/01/lense.pdf}

\enlargethispage{20pt}
\bibitem{Adler-Bazin-Schiffer}
Ronald J. Adler, Maurice Bazin, and Menahem Schiffer, \\
 {\sl Introduction to General Relativity}, Second edition, \\
 (McGraw--Hill, New York, 1975).\\
 {}[It is important to acquire the 1975 second edition, the 1965 first edition does not contain any discussion of the Kerr spacetime.]
 
 %=====================================
 
\bibitem{MTW}
Charles Misner, Kip Thorne, and John Archibald Wheeler,  {\sl Gravitation}, \\
(Freeman, San Francisco, 1973).

\bibitem{Wald}
Robert Wald,
\emph{General relativity},\\
(University of  Chicago Press, Chicago, 1984).

\bibitem{weinberg}
Steven Weinberg,\\
\emph{Gravitation and Cosmology: Principles and Applications of the General Theory of Relativity},
(Wiley,  Hoboken, 1972).

\enlargethispage{10pt}
\bibitem{Hobson}
M.~P.~Hobson, G.~P.~Estathiou, and A~N.~Lasenby,\\
\emph{General relativity: An introduction for physicists},\\
(Cambridge University Press, Cambridge, UK, 2006).

\bibitem{D'Inverno}
Ray D'Inverno, {\sl Introducing Einstein's Relativity}, \\
(Oxford University Press, Oxford, UK, 1992).

\bibitem{Hartle}
James Hartle, {\sl Gravity: An introduction to Einstein's general relativity},\\
(Addison Wesley, San Francisco, 2003).

\bibitem{Carroll}
Sean Carroll, {\sl  An introduction to general relativity: Spacetime and Geometry},
(Addison Wesley, San Francisco, 2004).

\bibitem{kerr-intro}
M.~Visser,
``The Kerr spacetime: A brief introduction'',
[\arXiv{0706.0622} [gr-qc]].
Published in \cite{kerr-book}.
%94 citations counted in INSPIRE as of 13 Jun 2020


%\clearpage
\bibitem{kerr-book}
D.~L.~Wiltshire, M.~Visser and S.~M.~Scott (editors),\\
\emph{The Kerr spacetime: Rotating black holes in general relativity},\\
(Cambridge University Press, Cambridge, 2009).
%31 citations counted in INSPIRE as of 13 Jun 2020

%======================================

\bibitem{Kerr}
Roy Kerr, \\
``Gravitational field of a spinning mass as an example of algebraically special metrics'',
 Physical Review Letters {\bf  11} 237-238 (1963).
 \doi{10.1103/PhysRevLett.11.237}

\bibitem{Kerr-Texas}
Roy Kerr,\\
``Gravitational collapse and rotation'',\\
published in: 
{\sl Quasi-stellar sources and gravitational collapse:
Including the proceedings of the First Texas Symposium on Relativistic
Astrophysics}, edited by Ivor Robinson, Alfred Schild, and E.L. Sch\"ucking
(University of Chicago Press, Chicago, 1965), pages 99--102.\\
The conference was held in Austin, Texas, on 16--18 December 1963.

\bibitem{kerr-newman}
E. Newman, E. Couch, K. Chinnapared, A. Exton, A. Prakash and R. Torrence,
``Metric of a Rotating, Charged Mass'',
J. Math. Phys. \textbf{6} (1965) 918,
\doi{10.1063/1.1704351}

\bibitem{race}
G. Dautcourt,
``Race for the Kerr field'',
Gen Relativ Gravit  {\bf41} (2009) 1437--1454,
\doi{10.1007/s10714-008-0700-y}

\clearpage
\bibitem{kerr-book-2}
Barrett O'Neill,
\emph{The geometry of Kerr black holes},\\
(Peters, Wellesley, 1995). Reprinted (Dover, Mineloa, 2014).
\\
ISBN-13: 978-0486493428, ISBN-10: 0486493423

%===================================== 

\bibitem{river}
A.~J.~Hamilton and J.~P.~Lisle,\\
``The River model of black holes'',\\
Am. J. Phys. \textbf{76} (2008), 519-532
\doi{10.1119/1.2830526}\\{}
[\arXiv{gr-qc/0411060} [gr-qc]].

\bibitem{Doran}
C.~Doran,
``A New form of the Kerr solution'',
Phys. Rev. D \textbf{61} (2000), 067503
\doi{10.1103/PhysRevD.61.067503}
[\arXiv{gr-qc/9910099} [gr-qc]].
%100 citations counted in INSPIRE as of 01 Nov 2021


\bibitem{Gordon-form}
S.~Liberati, G.~Tricella and M.~Visser,
``Towards a Gordon form of the Kerr spacetime'',
Class. Quant. Grav. \textbf{35} (2018) no.15, 155004
\doi{10.1088/1361-6382/aacb75}
[\arXiv{1803.03933} [gr-qc]].
%7 citations counted in INSPIRE as of 01 Nov 2021

%======================================
%======================================
 
 \bibitem{Carballo-Rubio:2018}
R.~Carballo-Rubio, F.~Di Filippo, S.~Liberati and M.~Visser,
``Phenomenological aspects of black holes beyond general relativity'',
Phys. Rev. D \textbf{98} (2018) no.12, 124009
\doi{10.1103/PhysRevD.98.124009}
[\arXiv{1809.08238} [gr-qc]].
%75 citations counted in INSPIRE as of 05 Jun 2021

\bibitem{Visser:2009a}
M.~Visser, C.~Barcel\'o, S.~Liberati and S.~Sonego,\\
``Small, dark, and heavy: But is it a black hole?'',\\
PoS \textbf{BHGRS} (2008), 010
\doi{10.22323/1.075.0010}
[\arXiv{0902.0346} [gr-qc]].
%54 citations counted in INSPIRE as of 05 Jun 2021

\bibitem{Visser:2009b}
M.~Visser,
``Black holes in general relativity'',
PoS \textbf{BHGRS} (2008), 001
\doi{10.22323/1.075.0001}
[\arXiv{0901.4365} [gr-qc]].
%38 citations counted in INSPIRE as of 05 Jun 2021

%========================================

\bibitem{Vincent:2012}
F.~H.~Vincent, E.~Gourgoulhon and J.~Novak,\\
``3+1 geodesic equation and images in numerical spacetimes'',\\
Class. Quant. Grav. \textbf{29} (2012), 245005
\doi{10.1088/0264-9381/29/24/245005}
[\arXiv{1208.3927} [gr-qc]].
%24 citations counted in INSPIRE as of 03 Dec 2021

\bibitem{Vincent:2020}
F.~H.~Vincent, M.~Wielgus, M.~A.~Abramowicz, E.~Gourgoulhon, J.~P.~Lasota, T.~Paumard and G.~Perrin,
\\
``Geometric modeling of M87* as a Kerr black hole or a non-Kerr compact object'',
\\
Astron. Astrophys. \textbf{646} (2021), A37
\doi{10.1051/0004-6361/202037787}
[\arXiv{2002.09226} [gr-qc]].
%28 citations counted in INSPIRE as of 03 Dec 2021

\bibitem{Bambi:2012}
C.~Bambi,
``A code to compute the emission of thin accretion disks in non-Kerr space-times and test the nature of black hole candidates'',\\
Astrophys. J. \textbf{761} (2012), 174
\doi{10.1088/0004-637X/761/2/174}\\{}
[\arXiv{1210.5679} [gr-qc]].
%109 citations counted in INSPIRE as of 03 Dec 2021

\bibitem{Glampedakis:2018}
K.~Glampedakis and G.~Pappas,
``Modification of photon trapping orbits as a diagnostic of non-Kerr spacetimes'',
Phys. Rev. D \textbf{99} (2019) no.12, 124041
\doi{10.1103/PhysRevD.99.124041}
[\arXiv{1806.09333} [gr-qc]].
%23 citations counted in INSPIRE as of 03 Dec 2021

%======================================

\clearpage
\bibitem{Kerr-Darboux}
Joshua Baines, Thomas Berry, Alex Simpson, and Matt Visser, \\
``Darboux diagonalization of the spatial 3-metric in Kerr spacetime",\\
  Gen.Rel.Grav. \textbf{53} (2021) 1, 3 \doi{10.1007/s10714-020-02765-0}
  \\{} [\arXiv{2009.01397} [gr-qc]]
  
 \bibitem{Valiente-Kroon:2004a}
 J.~A.~Valiente Kroon,
``On the nonexistence of conformally flat slices in the Kerr and other stationary space-times'',
Phys. Rev. Lett. \textbf{92} (2004), 041101
\doi{10.1103/PhysRevLett.92.041101}
[\arXiv{gr-qc/0310048} [gr-qc]].
%36 citations counted in INSPIRE as of 29 May 2021

\bibitem{Valiente-Kroon:2004b}
J.~A.~Valiente Kroon,
``Asymptotic expansions of the Cotton-York tensor on slices of stationary space-times'',
Class. Quant. Grav. \textbf{21} (2004), 3237-3250
\doi{10.1088/0264-9381/21/13/009}
[\arXiv{gr-qc/0402033} [gr-qc]].
%29 citations counted in INSPIRE as of 30 May 2021
 
 \bibitem{Jaramillo:2007}
J.~L.~Jaramillo, J.~A.~Valiente Kroon and E.~Gourgoulhon,
``From geometry to numerics: Interdisciplinary aspects in mathematical and numerical relativity'',\\
Class. Quant. Grav. \textbf{25} (2008), 093001
\doi{10.1088/0264-9381/25/9/093001}
[\arXiv{0712.2332} [gr-qc]].
%18 citations counted in INSPIRE as of 30 May 2021

%=========================================
\bibitem{Hioki:2009}
K.~Hioki and K.~i.~Maeda,
``Measurement of the Kerr Spin Parameter by Observation of a Compact Object's Shadow'',
Phys. Rev. D \textbf{80} (2009), 024042
\doi{10.1103/PhysRevD.80.024042}
[\arXiv{0904.3575} [astro-ph.HE]].
%250 citations counted in INSPIRE as of 03 Dec 2021

\bibitem{Wilkins:1972}
D.~C.~Wilkins,
``Bound Geodesics in the Kerr Metric'',
Phys. Rev. D \textbf{5} (1972), 814-822
\doi{10.1103/PhysRevD.5.814}
%172 citations counted in INSPIRE as of 03 Dec 2021

\bibitem{Page:2006}
D.~N.~Page, D.~Kubiznak, M.~Vasudevan and P.~Krtous,
``Complete integrability of geodesic motion in general Kerr-NUT-AdS spacetimes'',
Phys. Rev. Lett. \textbf{98} (2007), 061102
\doi{10.1103/PhysRevLett.98.061102}
[\arXiv{hep-th/0611083} [hep-th]].
%122 citations counted in INSPIRE as of 03 Dec 2021

\bibitem{Pretorius:2007}
F.~Pretorius and D.~Khurana,
``Black hole mergers and unstable circular orbits'',
Class. Quant. Grav. \textbf{24} (2007), S83-S108
\doi{10.1088/0264-9381/24/12/S07}
[\arXiv{gr-qc/0702084} [gr-qc]].
%101 citations counted in INSPIRE as of 03 Dec 2021


\bibitem{Fujita:2009}
R.~Fujita and W.~Hikida,
``Analytical solutions of bound timelike geodesic orbits in Kerr spacetime'',
Class. Quant. Grav. \textbf{26} (2009), 135002
\doi{10.1088/0264-9381/26/13/135002}
[\arXiv{0906.1420} [gr-qc]].
%98 citations counted in INSPIRE as of 03 Dec 2021

  
 \bibitem{Hackmann:2010}
E.~Hackmann, C.~Lammerzahl, V.~Kagramanova and J.~Kunz,\\
``Analytical solution of the geodesic equation in Kerr-(anti) de Sitter space-times'',\\
Phys. Rev. D \textbf{81} (2010), 044020
\doi{10.1103/PhysRevD.81.044020}\\{}
[\arXiv{1009.6117} [gr-qc]].
%101 citations counted in INSPIRE as of 03 Dec 2021
 
 \bibitem{Sereno:2006}
M.~Sereno and F.~De Luca,\\
``Analytical Kerr black hole lensing in the weak deflection limit'',\\
Phys. Rev. D \textbf{74} (2006), 123009
\doi{10.1103/PhysRevD.74.123009}
[\arXiv{astro-ph/0609435} [astro-ph]].
%45 citations counted in INSPIRE as of 03 Dec 2021

\enlargethispage{20pt}
\bibitem{Sereno:2007}
M.~Sereno and F.~De Luca,\\
``Primary caustics and critical points behind a Kerr black hole'',\\
Phys. Rev. D \textbf{78} (2008), 023008
\doi{10.1103/PhysRevD.78.023008}\\{}
[\arXiv{0710.5923} [astro-ph]].
%29 citations counted in INSPIRE as of 03 Dec 2021

 \bibitem{Gralla:2019-a}
S.~E.~Gralla and A.~Lupsasca,
``Lensing by Kerr Black Holes'',\\
Phys. Rev. D \textbf{101} (2020) no.4, 044031
\doi{10.1103/PhysRevD.101.044031}\\{}
[\arXiv{1910.12873} [gr-qc]].
%43 citations counted in INSPIRE as of 03 Dec 2021

\bibitem{Gralla:2019-b}
S.~E.~Gralla and A.~Lupsasca,\\
``Null geodesics of the Kerr exterior'',
Phys. Rev. D \textbf{101} (2020) no.4, 044032
\doi{10.1103/PhysRevD.101.044032}
[\arXiv{1910.12881} [gr-qc]].
%27 citations counted in INSPIRE as of 03 Dec 2021

\bibitem{Warburton:2013}
N.~Warburton, L.~Barack and N.~Sago,
``Isofrequency pairing of geodesic orbits in Kerr geometry'',
Phys. Rev. D \textbf{87} (2013) no.8, 084012
\doi{10.1103/PhysRevD.87.084012}
[\arXiv{1301.3918} [gr-qc]].
%33 citations counted in INSPIRE as of 03 Dec 2021 
 
 \bibitem{deFelice:1968}
F.~de Felice,
``Equatorial geodesic motion in the gravitational field of a rotating source'',
Nuovo Cim. B \textbf{57} (1968), 351
\doi{10.1007/BF02710207}
%31 citations counted in INSPIRE as of 03 Dec 2021

\bibitem{Rana:2019}
P.~Rana and A.~Mangalam,
``Astrophysically relevant bound trajectories around a Kerr black hole'',
Class. Quant. Grav. \textbf{36} (2019), 045009
\doi{10.1088/1361-6382/ab004c}
[\arXiv{1901.02730} [gr-qc]].
%18 citations counted in INSPIRE as of 03 Dec 2021

\bibitem{Gariel:2007}
J.~Gariel, M.~A.~H.~MacCallum, G.~Marcilhacy and N.~O.~Santos,\\
``Kerr Geodesics, the Penrose Process and Jet Collimation by a Black Hole'',\\{}
[\arXiv{gr-qc/0702123} [gr-qc]].
%7 citations counted in INSPIRE as of 03 Dec 2021

\bibitem{Gariel:2016}
J.~Gariel, N.~O.~Santos and A.~Wang,\\
``Observable acceleration of jets by a Kerr black hole'',\\
Gen. Rel. Grav. \textbf{49} (2017) no.3, 43
\doi{10.1007/s10714-017-2208-9}\\{}
[\arXiv{1610.01241} [gr-qc]].
%7 citations counted in INSPIRE as of 03 Dec 2021
 
\bibitem{Paganini:2016}
C.~F.~Paganini, B.~Ruba and M.~A.~Oancea,
``Characterization of Null Geodesics on Kerr Spacetimes''.
[\arXiv{1611.06927} [gr-qc]].
%7 citations counted in INSPIRE as of 03 Dec 2021
  
\bibitem{Lammerzahl:2015}
C.~L\"ammerzahl and E.~Hackmann,
``Analytical Solutions for Geodesic Equation in Black Hole Spacetimes'',
Springer Proc. Phys. \textbf{170} (2016), 43-51
\doi{10.1007/978-3-319-20046-0\_5}
[\arXiv{1506.01572} [gr-qc]].
%4 citations counted in INSPIRE as of 03 Dec 2021  

\bibitem{Boccaletti:2005}
Dino Boccaletti, Francesco Catoni, Roberto Cannata, Paolo Zampetti,\\
``Integrating the geodesic equations in the Schwarzschild and Kerr space-times using Beltrami's geometrical method'',
  General Relativity and Gravitation, 37(12) (2005) 2261
\doi{10.1007/s10714-005-0203-z}
[\arXiv{gr-qc/0502051} [gr-qc]].
  
%=========================================
 
 \bibitem{LISA} 
E.~Barausse, E.~Berti, T.~Hertog, S.~A.~Hughes, P.~Jetzer, P.~Pani, T.~P.~Sotiriou, N.~Tamanini, H.~Witek and K.~Yagi, \textit{et al.}
``Prospects for Fundamental Physics with LISA'',
Gen. Rel. Grav. \textbf{52} (2020) no.8, 81
\doi{10.1007/s10714-020-02691-1}
[\arXiv{2001.09793} [gr-qc]].
%96 citations counted in INSPIRE as of 01 Nov 2021

\bibitem{Echos}
V.~Cardoso, S.~Hopper, C.~F.~B.~Macedo, C.~Palenzuela and P.~Pani,
``Gravitational-wave signatures of exotic compact objects and of quantum corrections at the horizon scale'',
Phys. Rev. D \textbf{94} (2016) no.8, 084031
\doi{10.1103/PhysRevD.94.084031}
[\arXiv{1608.08637} [gr-qc]].
%293 citations counted in INSPIRE as of 01 Nov 2021

\clearpage
\bibitem{bb-Kerr}
J.~Mazza, E.~Franzin and S.~Liberati,
``A novel family of rotating black hole mimickers'',
JCAP \textbf{04} (2021), 082
\doi{10.1088/1475-7516/2021/04/082}
[\arXiv{2102.01105} [gr-qc]].
%10 citations counted in INSPIRE as of 05 Jun 2021

\bibitem{bb-KN}
E.~Franzin, S.~Liberati, J.~Mazza, A.~Simpson and M.~Visser,
``Charged black-bounce spacetimes'',
JCAP \textbf{07} (2021), 036
\doi{10.1088/1475-7516/2021/07/036}
[\arXiv{2104.11376} [gr-qc]].
%0 citations counted in INSPIRE as of 30 May 2021

\bibitem{eye-of-storm}
A.~Simpson and M.~Visser,
``The eye of the storm: A regular Kerr black hole'',
[\arXiv{2111.12329} [gr-qc]].
%0 citations counted in INSPIRE as of 04 Dec 2021

%========================================

\bibitem{Finn} 
F.~Gray and D.~Kubiz\v{n}\'ak,
``Slowly rotating black holes with exact Killing tensor symmetries'',
[\arXiv{2110.14671} [gr-qc]].
%0 citations counted in INSPIRE as of 01 Nov 2021
 %=========================================================
  
\bibitem{Papadopoulos:2020}
G.~O.~Papadopoulos and K.~D.~Kokkotas,\\
``On Kerr black hole deformations admitting a Carter constant and an invariant criterion for the separability of the wave equation'',\\
Gen. Rel. Grav. \textbf{53} (2021) no.2, 21
\doi{10.1007/s10714-021-02795-2}
{}[\arXiv{2007.12125} [gr-qc]].
%0 citations counted in INSPIRE as of 15 Sep 2020

\bibitem{Papadopoulos:2018}
G.~O.~Papadopoulos and K.~D.~Kokkotas,\\
``Preserving Kerr symmetries in deformed spacetimes'',\\
Class. Quant. Grav. \textbf{35} (2018) no.18, 185014
\doi{10.1088/1361-6382/aad7f4}\\{} [\arXiv{1807.08594} [gr-qc]].
%12 citations counted in INSPIRE as of 15 Sep 2020


\bibitem{Benenti:1979}
S. Benenti and M. Francaviglia,
``Remarks on Certain Separability Structures and Their Applications to General Relativity'', \\
General Relativity and Gravitation {\bf10} (1979) 79--92,
\doi{10.1007/BF00757025}

%======================================


\bibitem{Frolov:2017}
V.~Frolov, P.~Krtous and D.~Kubiznak,\\
``Black holes, hidden symmetries, and complete integrability'',\\
Living Rev. Rel. \textbf{20} (2017) no.1, 6
\doi{10.1007/s41114-017-0009-9}\\{}
[\arXiv{1705.05482} [gr-qc]].
%83 citations counted in INSPIRE as of 16 Apr 2021

\bibitem{Giorgi}
E.~Giorgi,
``The Carter tensor and the physical-space analysis in perturbations of Kerr-Newman spacetime'',
[\arXiv{2105.14379} [gr-qc]].
%0 citations counted in INSPIRE as of 05 Jun 2021



%==================================
%==================================
%==================================
%\clearpage

\bibitem{hyper}
Ultra-elliptic and hyper-elliptic integrals, 
\url{https://encyclopediaofmath.org/wiki/Hyper-elliptic_integral}

\bibitem{hyper2}
Jeroen Spandaw and Duco van Straten,
``Hyperelliptic integrals and generalized arithmetic-geometric mean'',
Ramanujan Journal {\bf28}  (2012) 61--78 \doi{10.1007/s11139-011-9353-7}

\bibitem{elliptic}
Elliptic integrals, 
\url{https://mathworld.wolfram.com/EllipticIntegral.html}
\url{https://mathworld.wolfram.com/CarlsonEllipticIntegrals.html}

\bibitem{abelian}
Abelian integrals, \blue{\url{https://encyclopediaofmath.org/wiki/Abelian_integral}}

%===================================

\bibitem{Yang:2013-a}
X.~Yang and J.~Wang,
``YNOGK: A new public code for calculating null geodesics in the Kerr spacetime'',
Astrophys. J. Suppl. \textbf{207} (2013), 6
\doi{10.1088/0067-0049/207/1/6}
[\arXiv{1305.1250} [astro-ph.HE]].
%17 citations counted in INSPIRE as of 03 Dec 2021

\bibitem{Yang:2013-b}
X.~L.~Yang and J.~C.~Wang,
``YNOGKM: A new public code for calculating time-like geodesics in the Kerr-Newman spacetime'',
Astron. Astrophys. \textbf{561} (2014), A127
\doi{10.1051/0004-6361/201322565}
[\arXiv{1311.4436} [gr-qc]].
%8 citations counted in INSPIRE as of 03 Dec 2021

\bibitem{Chan:2017}
C.~k.~Chan, L.~Medeiros, F.~Ozel and D.~Psaltis,
``GRay2: A General Purpose Geodesic Integrator for Kerr Spacetimes'',
Astrophys. J. \textbf{867} (2018) no.1, 59
\doi{10.3847/1538-4357/aadfe5}
[\arXiv{1706.07062} [astro-ph.HE]].
%9 citations counted in INSPIRE as of 03 Dec 2021


%\bigskip
%\hrule
%\bigskip

%%===================================
%%===================================
%%===================================
%%===================================
%%===================================
\end{thebibliography}
\end{document}